\documentclass[12pt]{article}
\usepackage[utf8]{inputenc}
\usepackage[DIV=13]{typearea}
\usepackage{ragged2e}
\usepackage{calligra,amsmath,amsfonts,bbm,mathrsfs,amssymb,mathtools}
\usepackage{slashed,cancel}
\usepackage{cite}
\usepackage{bm} 
\usepackage{hyperref}
\hypersetup{colorlinks=true,urlcolor=blue,anchorcolor=blue,citecolor=blue,filecolor=blue,
            linkcolor=blue,menucolor=blue, linktocpage=true,pdfproducer=medialab}
\usepackage[english]{babel}
\usepackage{indentfirst}
\usepackage{graphicx}
\usepackage{float}
\usepackage{enumerate}
\usepackage{caption}
\usepackage[dvipsnames]{xcolor}
\usepackage{subfigure}
\usepackage{multirow}
\usepackage[normalem]{ulem} 
\usepackage{booktabs}
\usepackage{physics}
\usepackage{MnSymbol}
\numberwithin{equation}{section}

\definecolor{rosso}{cmyk}{0,1,1,0.4}
\definecolor{rossos}{cmyk}{0,1,1,0.55}
\definecolor{rossoc}{cmyk}{0,1,1,0.2}
\definecolor{blu}{cmyk}{1,1,0,0.3}
\definecolor{verde}{cmyk}{0.92,0,0.59,0.25}

\definecolor{pastelred}{rgb}{1,0.3333,0.3333}
\definecolor{pastelblue}{rgb}{0.3333,0.3333,1}
\definecolor{darkred}{rgb}{0.6666,0,0}
\definecolor{darkblue}{rgb}{0,0,0.6666}
\definecolor{brown}{rgb}{0.6,0.4,0.2}
\definecolor{purple}{rgb}{0.5,0,0.5}

\definecolor{myu}{rgb}{1,0,0}
\definecolor{myd}{rgb}{0,0.749,0.0784}
\definecolor{myt}{rgb}{0,0.392,0.9647}
\definecolor{myf}{rgb}{0.60, 0.20, 0.80}

\definecolor{myred}{rgb}{1,0,0}
\definecolor{myyellow}{rgb}{0.9294,0.69411,0.12549}
\definecolor{myblue}{rgb}{0,0.392,0.9647}

\definecolor{redG}{rgb}{0.85, 0.10, 0.10}
\definecolor{orangeG}{rgb}{1.00, 0.50, 0.00}
\definecolor{purpleG}{rgb}{0.60, 0.20, 0.80}

\newcommand{\be}{\begin{equation}}
\newcommand{\ee}{\end{equation}}
\newcommand{\ba}{\begin{equation}\begin{aligned}}
\newcommand{\ea}{\end{aligned}\end{equation}}

\newcommand{\ov}[1]{\overline{#1}}

\newcommand{\eV}{\,\rm{eV}}

\newcommand{\GeV}{\,\rm{GeV}}

\def\Re{{\texttt{Re}}}

\newcommand\subsetsim{\mathrel{%
  \ooalign{\raise0.2ex\hbox{$\subset$}\cr\hidewidth\raise-0.8ex\hbox{\scalebox{0.9}{$\sim$}}\hidewidth\cr}}}
\newcommand*{\diff}[1]{\text{d}#1}

\newcommand{\myu}[1]{\emph{\color{myu}{\bf #1}}}
\newcommand{\myd}[1]{\emph{\color{myd}{\bf #1}}}
\newcommand{\myt}[1]{\emph{\color{myt}{\bf #1}}}
\newcommand{\myf}[1]{\emph{\color{myf}{\bf #1}}}

\newcommand{\redG}[1]{\emph{\color{redG}{\bf #1}}}
\newcommand{\orangeG}[1]{\emph{\color{orangeG}{\bf #1}}}
\newcommand{\purpleG}[1]{\emph{\color{purpleG}{\bf #1}}}

\graphicspath{{Draft_Figures/}}

\begin{document} 
\renewcommand*{\thefootnote}{\fnsymbol{footnote}}

\begin{titlepage}

\vspace*{-1cm}
\flushleft{IFT-UAM/CSIC-24-122} 
\\[1cm]
\vskip 1cm

\begin{center}
{\bf \LARGE Star Shearing Season} \\[0.2cm]{\bf \LARGE --} \\[0.2cm]
{\bf \large Transient Signals in Wave-like Dark Matter Experiments from Black Hole Formation}
\centering
\vskip .3cm
\end{center}
\vskip 0.5  cm
\begin{center}
{\large\bf Arturo de Giorgi$^{a}$ and Joerg Jaeckel$^{b}$ },
\vskip .7cm
{\footnotesize
$^a$ Departamento de Fisica Teorica and Instituto de Fisica Teorica UAM/CSIC, Universidad Autonoma de Madrid, 
Cantoblanco, 28049, Madrid, Spain\\
$^b$Institut f\"ur Theoretische Physik, Universit\"at 
Heidelberg, Philosophenweg 16, 69120 Heidelberg, Germany\\
{\par\centering \vskip 0.25 cm\par}
}
\end{center}
\vskip 2cm
\begin{abstract}
\justify
Ordinary matter coupled to light weakly interacting bosons can lead to the formation of a macroscopic bosonic field in the vicinity of large matter concentrations such as ordinary or neutron stars. When these objects are turned into black holes due to a supernova or a binary merger this ``hair'' could be ``shorn'' off. Part of the field configuration would then be released leading to an outgoing field wave. For small masses this field transient remains rather compact and can induce a transient signal in experiments, in particular those that look for wave-like dark matter. This signal can be correlated with the corresponding astrophysical signal of the event. In this note, we consider a variety of couplings and the associated signals and estimate the corresponding sensitivities.
\end{abstract}
\end{titlepage}
\setcounter{footnote}{0}

\newpage
\pdfbookmark[1]{Table of Contents}{tableofcontents}
\tableofcontents

\renewcommand*{\thefootnote}{\arabic{footnote}}
%
%
%

\newpage

\section{Introduction}

In the search for ultralight bosons a current focus is on the possibility that they are dark matter~\cite{Preskill:1982cy,Abbott:1982af,Dine:1982ah,Piazza:2010ye,Nelson:2011sf,Arias:2012az}.  This has sparked a large variety of creative and novel search strategies (cf., e.g.~\cite{Adams:2022pbo,Antypas:2022asj} for recent reviews).
Most of these are focused on oscillating signals with a large coherence time. For very light bosonic dark matter this is expected, since at Earth's location the typical velocity of the dark matter particles is of the order of $v\sim 10^{-3}$ and this also sets the spread in the velocity spectrum $\Delta v\sim 10^{-3}$ as well as the corresponding spread in energies $\Delta E\sim 10^{-6} m$, where $m$ is the mass of the boson in question.
Accordingly, most experiments are focused on this type of signal.

\begin{figure}
    \centering
    \includegraphics[width=\textwidth]{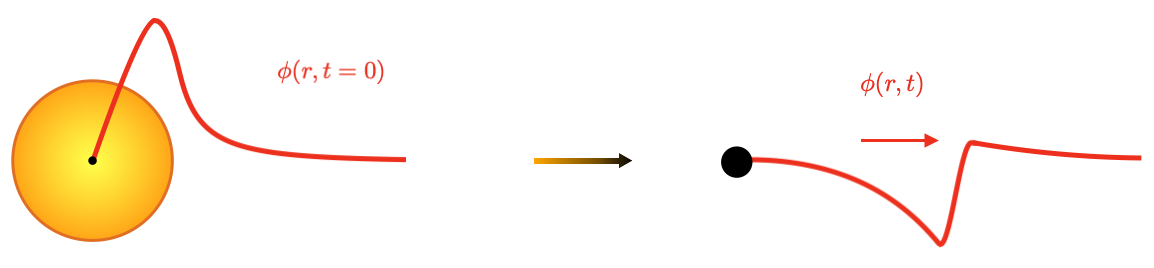}
    \caption{Cartoon of the initial static field $\phi(r,t=0)$ and its propagation after the source has collapsed into a black hole. The energy initially stored in the source propagates in space in a tsunami-like manner. The steep field gradient can be detected on Earth at different experimental facilities.}
    \label{fig:cartoon}
\end{figure}

However, these experiments may also have interesting sensitivity to other, in particular transient, signals. This is what we will pursue in this paper. In particular, we will consider the possibility that such a signal is generated when large amounts of ordinary matter are converted into a black hole. This could be when a star ends its life in a supernova (SN) explosion or in a situation where a neutron star in a binary with a black hole or another neutron star merges into a black hole.
While here we only pursue the phenomenology and leave a detailed theoretical description to future work, we have the following very simplistic, qualitative physical picture in mind, cf.~Fig.~\ref{fig:cartoon}. Let us consider a light scalar field linearly coupled to matter, e.g. baryons. Before the SN explosion, the matter density sources an extended Yukawa-type field around the star. During a SN explosion, a significant fraction of the baryons ends in the black hole and therefore cannot source the bosonic field anymore (another part is ``blown away''). The corresponding part of the field outside of the BH now does not have a source anymore (black holes having no hair~\cite{Israel:1967wq,Israel:1967za,Carter:1971zc,Ruffini:1971bza,Carter:2009nex,Robinson:1975bv,Carter:1979wef,Mazur:2000pn}) and may be (partially) released. While a certain fraction (perhaps roughly half) ends in the black hole, the remainder could propagate outside leading to a sizable field wave propagating outwards. Upon arriving at Earth this can then induce a transient signal in experiments searching for light bosons. This signal is directly correlated in time to other observable effects of the SN (e.g. light, or neutrinos) or merger event (e.g. gravitational waves). This provides a clear signature and can be used to discriminate from backgrounds. Moreover, as we will also detail in the following, the expected frequency spectrum of the signal is linked to the size of the progenitor object, thereby providing another interesting feature. Similar effects would not only be expected from linear couplings but also from other types of coupling (see, e.g.~\cite{Hook:2017psm,Balkin:2021zfd,Balkin:2023xtr} for examples in the case of axion-like particles). We can also imagine that the field (or at least a sizable fraction of its energy) is concentrated in a localized scalar field configuration linked to the existence of a dense clump of matter coupled to the light bosonic field, instead of having a long-range $\sim 1/r$ behaviour. Remarkably, the signal's shape (power spectrum) can clearly distinguish between Yukawa-like and compact sources. 
At this point we would like to stress the limitations of our simplistic modelling. A more definite prediction of the signal will need to take into account the dynamics of the collapse and the resulting evolution of the scalar field as well as gravitational effects, including the amount of field energy absorbed into the black hole during formation as well as that remaining bound to it (cf.~\cite{Barranco:2012qs,Cardoso:2022nzc} for some work in this direction).\footnote{We would like to thank the anonymous referee for noting these points.} Importantly this can include features such as the expected range of frequencies. We will comment on this in a bit more detail in Sec.~\ref{subsection:modeling}, but more detailed calculations are beyond the scope of the present work. In this light, our simplistic modelling and the signals derived from it should be taken only as illustrative of the range of possibilities. 

Previous work has, of course, also considered interesting transient signals. For example, such signals could also result from the passage of domain walls (that may form part of the dark matter) or other localized structures through GPS satellites or magnetometers~\cite{Pospelov:2012mt,Jaeckel:2016jlh,Roberts:2017hla,Roberts:2018xqn,Afach:2021pfd,GNOME:2023rpz,Cuadrat-Grzybowski:2024uph}. Also bosenova-type events, possibly resulting from a merger of boson stars or accretion of extra dark matter onto a boson star could lead to specific types of transient features~\cite{Eby:2021ece,Arakawa:2023gyq,Maseizik:2024qly}, detectable in Earth bound experiments~\cite{Gramolin:2020ict,Salemi:2021gck,DMRadio:2022pkf}. 
Signals correlated with a cosmic event that is detectable by other means could also arise from two black holes or two neutron stars, featuring a scalar cloud, merging~\cite{Krause:1994ar,Barausse:2012da,Dailey:2020sxa,GNOME:2023rpz}.  As in the case we envision, this would feature a timewise coincidence with the gravitational wave signal from the merger that can aid the detection~\cite{Dailey:2020sxa,GNOME:2023rpz}.

Our investigation expands this class of interesting signals and adds new types of sources at different length scales and coincident signals. More specifically, we consider signals that originate from an (assumed) abrupt ``vanishing'' of the coupling of the bosonic field to the matter object when a black hole is formed. This leads to two main features. First, such a signal is also possible for a single object, e.g. a star being converted into a black hole in a supernova event. We therefore should also look out for coincidences with events such as supernovae, observable with both light and neutrinos. Second, the frequencies and the signal envelopes are determined by the size of the initial object rather than the merging process (e.g. considered in~\cite{Dailey:2020sxa,GNOME:2023rpz}), thereby opening different frequency and shape windows.
Moreover, we also consider a wide range of interesting experiments, including magnetometers~\cite{GNOME:2023rpz}, but also experiments such as LISA pathfinder~\cite{Frerick:2023xnf}, torsion pendulums~\cite{Shaw:2021gnp, Sun:2024qis} and an example of an axion haloscope~\cite{DMRadio:2022pkf}.

Our investigation proceeds along the following lines. In Sect.~\ref{sec:transient-signals} we discuss the signal and review its propagation to Earth, considering in particular its frequency spectrum and effects of a non-vanishing boson mass. The different experimental sensitivities are discussed in Sect.~\ref{sec:exp-sens}. We briefly summarize and conclude in Sect.~\ref{sec:conclusions}. Some additional technical details are provided in the appendix.

\section{Transient Signals}
\label{sec:transient-signals}
The starting point of our investigation is the presence of a static field configuration sourced by the presence of a macroscopic object such as a (neutron) star. 
In a very simplistic approximation, we consider this object to vanish, releasing the field as if no source is present.

To put this picture into practice we consider, for simplicity, a real scalar field, $\phi$.
The initial static field configuration is abruptly released at time $t=0$. 
From then on its time evolution for $t>0$ is dictated by the free Klein-Gordon~(KG) Lagrangian 
\begin{equation}
\label{eq:Lagrangian-free}
    \mathcal{L}_\text{free}=\dfrac{1}{2}\partial^\mu \phi \partial_\mu \phi-\dfrac{1}{2}m_\phi^2\phi^2\,,
\end{equation}
and the corresponding homogeneous KG equation
\begin{equation}
\label{eq:eom}
    (\Box +m_\phi^2)\phi(x) = 0\,.
\end{equation}

Our problem then consists of two steps. First, we would like to model a couple of interesting initial ($t<0$) field configurations. We do this in the following subsection~\ref{subsec:initial}. We then turn to the time evolution which determines the signal as it arrives on Earth in subsection~\ref{subsec:timevol}.

\subsection{Sources and Initial Configurations}\label{subsec:initial}
We consider two initial sources and field configurations. In both cases, the source is characterized by a typical length $R$. In the following, we analyze them separately and report the main results. More details can be found in App.~\ref{app:Klein-Gordon}.

\subsubsection{Yukawa-like Source}
The first field configuration, hereafter referred to as \textit{Yukawa-like~(YL) source}, is generated by linearly coupling the scalar field to the ordinary matter in the initial object,
 \begin{equation}
\label{eq:Lagrangian-source}
    \mathcal{L}_\text{source}=\sum_{\psi}g_{\phi\psi\psi}\phi\bar{\psi}\psi\equiv\phi J(x)\,.
\end{equation}
In this equation, $\psi$ can be any Standard Model fermion present in the source object and $g_{\phi\psi\psi}$ is the corresponding coupling to the scalar $\phi$. 
The source term can be summarized in the macroscopic ``charge'' density,
\begin{equation}
J(x)=\sum_{\psi}g_{\phi\psi\psi}n_{\psi}.
\end{equation}
In writing this we have used that, in the non-relativistic approximation suitable for a static source with only particles (and no or few antiparticles) present, the particle number density is related to the fields as,
\begin{equation}
    n_{\psi}=\bar{\psi}\psi\approx\psi^{\dagger}\psi.
\end{equation}

Further simplifying the problem by assuming a homogeneous distribution within a spherical source of radius $R$ which vanishes at $t=0$, we set,
\begin{equation}
    J(x)=g_\text{YL}\dfrac{3}{4\pi R^3}\Theta(-t)\Theta(R-r)\,.
\end{equation}
Here, $r\equiv |\vec{x}|$ is the distance from the center of the source and $g_\text{YL}=V\times\sum_{\psi}g_{\phi\psi\psi}n_{\psi}$ is the effective Yukawa-like charge in the volume $V=4\pi R^3/3$ generating the static field. 

The solution is given by
\begin{equation}
\label{eq:field_R}
    \phi(r,t< 0)=\dfrac{3g_\text{YL}}{2\pi^2R^2r}\int\limits_0^\infty \diff{k}\,\dfrac{1}{\omega_k^2}\dfrac{\sin(kr)}{k}\cos(t\omega_k\,\Theta(t))\left[\dfrac{\sin(k R)}{kR}-\cos{(kR)}\right]\,,
\end{equation}
where $\omega_k\equiv\sqrt{k^2+m_\phi^2}$.
For the case $m_\phi=0$ the result can be computed in a closed form for all times. This is a good approximation as long as $m_\phi r\ll 1$. We will analyze the time-dependent solution in more detail in the next subsection.

For $t<0$ the time dependence drops out of the integral consistently with the ansatz of an initially static source and yields the usual Coulomb-type potential for a homogeneously charged sphere, which in the case $m_\phi=0$ reads
\begin{align}
 &     \phi_\text{YL}(r\leq R) =\dfrac{g_\text{YL}}{8\pi R}\left(3-\dfrac{r^2}{R^2}\right)\,,\\
&\phi_\text{YL}(r\geq R)=\dfrac{g_\text{YL}}{4\pi r}\,.
\end{align}

A useful way to quantify the strength of the field is to consider its total static energy. 
The Hamiltonian density of the field reads
\begin{equation}
\label{eq:app-energy}
    \mathcal{H}=\dfrac{1}{2}\left[(\dot{\phi})^2+(\vec{\nabla}\phi)^2+m_\phi^2\phi^2-2\phi J(x)\right]\,.
\end{equation}
The energy stored in the field in the massless case reads
\begin{equation}
\begin{split}
        E&=\int\diff^3x\,\mathcal{H}=-\dfrac{3g_\text{YL}^2}{20\pi R}\,.
        \end{split}
\end{equation}
The result for the massive case can be found in App.~\ref{app:Klein-Gordon}.

This allows us to trade the coupling strength in favour of the energy stored by the field and parametrize 
\begin{equation}
    g_\text{YL}=\pm\sqrt{\frac{20\pi|E|R}{3}}\,.
\end{equation}

To get an impression of the, potentially large, amount of energy stored in such field configurations let us consider a simple $(B-L)$-type coupling. Taking into account solely the coupling with neutrons\footnote{This is roughly what remains after balancing protons and electrons for charge neutrality. But for our order of magnitude estimate, we do not account for the composition.}, the effective YL source coupling, $g_\text{YL}$, can be approximately related to $g_{\phi\psi\psi}$ via
\begin{equation}
    g_\text{YL}\sim g_{\phi\psi\psi} \dfrac{M}{m_n}\,,
\end{equation}
where $M$ is the mass of the source and $m_n$ the neutron mass.
This is a rather crude estimate. We will comment in more detail on the specific couplings separately in each section for the sake of the comparison with the literature.

The energy then reads
\begin{eqnarray}
|E|&\sim&0.2 M_\odot\left(\frac{g_{\phi\psi\psi}}{10^{-16}}\right)^2\left(\frac{R_{\odot}}{R}\right)\left(\frac{M}{ M_{\odot}}\right)^2
\\\nonumber
&\sim&0.1 M_\odot\left(\frac{g_{\phi\psi\psi}}{10^{-16}}\right)^2\left(\frac{800 R_{\odot}}{R}\right)\left(\frac{M}{20 M_{\odot}}\right)^2
\\\nonumber
&\sim&1 M_\odot\left(\frac{g_{\phi\psi\psi}}{10^{-18}}\right)^2\left(\frac{10\,{\rm km}}{R}\right)\left(\frac{M}{M_{\odot}}\right)^2.
\end{eqnarray}

Therefore a simple constraint that the field energy should not be too large already constrains the coupling to be quite small. As a rough estimate, we show in Fig.~\ref{fig:coupling-bounds} the constraint of the neutron star as a purple line and lightly shaded area.

When discussing the sensitivities in Sec.~\ref{sec:exp-sens} we will use $E$ to parametrize the field strength. Thereby we do not necessarily insist on the field being generated by the same coupling as probed in the experiment on Earth.

\subsubsection{Compact-Source}
As a second option, we assume a field configuration where most of the energy is concentrated in a finite region around the star, i.e. having a sharper drop-off than $\sim 1/r$. 
The main purpose of this is phenomenological, i.e. to see how this affects the signal. Nevertheless, we also think that this can be useful to model part of configurations sourced by non-linear interactions as well as self-interactions, e.g.~\cite{Hook:2017psm,Balkin:2021zfd,Balkin:2023xtr}. That said, we point out that the latter also typically feature a $\sim 1/r$ component, as this is a solution of the static free vacuum equations. In this sense, a more general object is likely to have a combination of a compact part and some $\sim 1/r$ tail. In the following, we focus on modelling the compact part.

Sources of these types and the corresponding fields will be hereafter referred to as \textit{compact-sources} and \textit{compact-fields}, respectively. In this case, $R$ does not necessarily coincide with the star radius, but we naively expect it to be of the same order of magnitude. Because of the spherical symmetry, the function depends only on the distance from the source, $r\equiv |\Vec{x}|$. We take it to be non-zero in a finite domain
\begin{align}
    &\phi_\text{CS}(x)=\begin{cases}
        \phi_\text{CS}(r) & r\leq R\,,\\
        0 &\text{otherwise}\,,
    \end{cases}
\end{align}
and to be $C^2$-differentiable to simplify the discussion avoiding discontinuities in the KG equation with boundary condition
\begin{equation}
    \phi(R)=\phi'(R)=\phi''(R)=0\,.
\end{equation}
The simplest function that can be taken is a third-order polynomial of the form
\begin{equation}
\label{eq:CS-benchmark}
    \phi_\text{CS}(r) = \sqrt{\frac{35|E|R}{6\pi}} \frac{(R-r)^3}{R^4}\,,
\end{equation}
where $E$ is the energy stored in the field.
We will use it as a benchmark throughout this work. 

\subsubsection{Modelling limitations}\label{subsection:modeling}

Having described our simplistic modelling of the initial field configuration released upon black hole formation let us comment on the limitations of our approach.

A major limitation is neglecting the dynamics of the formation process as well as the effects of gravitational pull on the scalar field.\footnote{We again gratefully acknowledge the anonymous referee for highlighting them and in particular for their estimate of the relaxation time-scale.}
Beyond our approximation of the instantaneous disappearing of the source, when the black hole is formed, part of the scalar field configuration will be dragged along with the collapsing matter. This may significantly change the relevant size of the object, e.g. from the size of the initial star to something not much bigger than the final black hole. Moreover, part of the scalar field may be dragged within the event horizon and will not reach us. Moreover, some of the field may also remain bound in the vicinity of the final black hole (see~\cite{Barranco:2012qs,Cardoso:2022nzc}) and thereby not contribute to the signal that we aim to detect.

Our modelling effectively assumes that the collapse and therefore the vanishing of the source of the field configuration is faster than the scalar field's relaxation time. A sizable component of the scalar field may then be able to escape from the source before being swallowed. In this case, our modelling should be reasonable for both the Yukawa and the compact sources. 
Yet, for the nearly massless scalar fields we are considering, the propagation of signals and therefore relaxation may occur with nearly the speed of light, and therefore faster than the collapse of the progenitor.
So we also need to consider this possibility and we will do so momentarily. Let us nevertheless note that in particular, the compact source case may arise in the presence of significant self-interactions which will likely also affect the relaxation of the scalar field.

If the scalar field's relaxation time to its stable configuration is much faster than the collapse rate of the source into a black hole, then the scalar field is dragged in and all the energy stored in the source is swallowed. This implies no signal for the compact source case. The same does not necessarily apply to the Yukawa case as part of the energy is stored in the gradient of the long-range Coloumb tail. A particle outside the event horizon at some distance $r$ will be able to escape if it has a velocity larger than the escape one
\begin{equation}
      v_e=\sqrt{\frac{R_s}{r}}\,,
\end{equation}
where $R_s\sim R_{\rm Schwarzschild}= 2G_N M$ is the shrunken radius just before the black hole formation.
This is typically the case in our setup as our scalar field is ultralight.
That said, a full estimation of such effects is complex and requires detailed and involved numerical simulations within the frame of General Relativity, which is beyond the scope of this work.  

All in all, in the most optimistic scenario one would expect the same type of signal but rescaled taking into account the effective energy released in the process, $E_\text{tail}=\chi |E|$, where $\chi$ is the rescaling factor. 
First, the source adiabatically shrinks before turning into a black hole, effectively changing its energy content from $E$ to $E_s$
\begin{equation}
    E_s=\left(\frac{R}{R_s}\right)\,E\,.
\end{equation}
In this case, the condition $E\leq 0.1 M_\odot$ should be recast as $E_s\leq 0.1 M_\odot$, thus $E\leq 0.1(R_s/R)\, M_\odot $, which significantly weakens our constraints (exception for the Neutron Star case where $R\sim R_s$, see Sec.~\ref{sec:sources}).
Finally, the energy stored in the Coulomb tail that propagates outwards reads (cfr.~Eq.~\eqref{eq:energy-source})
\begin{eqnarray}
    E_\text{tail}=\frac{1}{2}\int\limits_{|\vec{x}|\geq R_s} d^3x\,\mathcal{H}=\frac{1}{4}\int\limits_{|\vec{x}|\geq R_s} d^3x\,(\phi')^2=\frac{g_\text{YL}^2}{16\pi R_s}=-\frac{5}{12}\left(\frac{R}{R_s}\right)E=-\frac{5}{12}E_s\,,
\end{eqnarray}
where the $1/2$ factor takes into account the outgoing half of the signal, leading to $\chi = 5/12$.
All in all, this would amount to rescaling the bounds in the plots of the following sections (summary in Fig.~\ref{fig:coupling-bounds}) by a factor
\begin{equation}
    g_\text{max} \to g_\text{max}\times \sqrt{\frac{12R}{5R_s}}\approx 0.9\,g_\text{max}\times \sqrt{\left(\frac{R}{10\,\text{Km}}\right)\left(\frac{10\,M_\odot}{M}\right)}\,.
\end{equation}
As can be seen from Tab.~\ref{tab:sources}, the rescaling does not dramatically affect the Neutron Star case, while it weakens the bounds derived from the Red Giant or the supernova by about four orders of magnitude.

\bigskip

In the following section, we will study how the field profiles evolve over time.

\subsection{Time Evolution}\label{subsec:timevol}
The evolution of the initial profile can be obtained by employing the KG equation. For simplicity, we give most of the formulas only for the massless case. However, we comment on the massive case and discuss the resulting limitations in the mass reach of Earth bound detectors.

Due to causality, the signal can arrive only after $t_\text{min}=r-R$ and last for a time $\Delta t \geq 2R$, where the equality holds in the massless case.
For a spherically symmetric source, the time evolution of the field for $t>0$ reads~\footnote{These integrals typically need the introduction of regulators at infinity, e.g. $e^{-\epsilon y}$ with $\epsilon>0$. Their presence will be omitted in the following.} (see App.~\ref{app:Klein-Gordon} or Ref.~\cite{Dailey:2020sxa} for application to spherically symmetric Gaussian ansatz)
\begin{equation}
    \phi(r,t)=\frac{4}{(2\pi)r}\int\limits_0^\infty \diff{k}\cos{(\omega_k t)}\sin(kr)\int\limits_{0}^\infty \diff{y}\, y\phi(y)\sin(ky)\,,
\end{equation}
where we use the notation $\phi(r)\equiv \phi(r,t=0)$.
For a massless field, the integrals can be explicitly calculated and read
\begin{equation}
    \begin{split}
        \phi(r,t)
        &=\dfrac{1}{2r}\left\{(r+t)\phi(r+t)+(r-t)\phi(r-t)\Theta(r-t)-(t-r)\phi(t-r)\Theta(t-r)\right\}\,,
    \end{split}
\end{equation}
and thus
\begin{align}
    &\label{eq:1-time}\phi(r,t<r)=\dfrac{1}{2r}\left[(r+t)\phi(r+t)+(r-t)\phi(r-t)\right]=\dfrac{u(r+t)+u(r-t)}{2r}\,,\\
    &\label{eq:2-time}\phi(r,t>r)=\dfrac{1}{2r}\left[(r+t)\phi(r+t)-(t-r)\phi(t-r)\right]=\dfrac{u(r+t)-u(t-r)}{2r}\,,
\end{align}
where we have introduced the parameterization $\phi(r)=u(r)/r$ to make the $1/r$ dependence at large field values more obvious.
For example, in the case of a CS, the initial field configuration vanishes for $r\geq R$ and thus $\phi_\text{CS}(r+t)=0$ $\forall t >0$ and $r\geq R$. 
Moreover, the signal is non-zero only for $t\in[r-R,r+R]$ which explicitly shows its duration to be $\Delta t = 2R$. 
In this region the argument of the function $u$ is always $|r-t|\leq R$ and thus at large $r$ the behaviour is dominated by the explicit factor $1/r$ on the right-hand sides of the above equations.

We also note that, independently of the specific initial configuration, all massless CS fields have zero transient-time average, i.e. $\left<\phi_\text{CS}(r,t)\right>_{\Delta t}=0$, see appendix~\ref{app:average}. Based on this common feature we focus on the benchmark proposed in Eq.~\eqref{eq:CS-benchmark}.

\begin{figure}[H]
    \centering
    \subfigure[\label{fig:field-evolution-1}]{\includegraphics[width=0.85\textwidth]{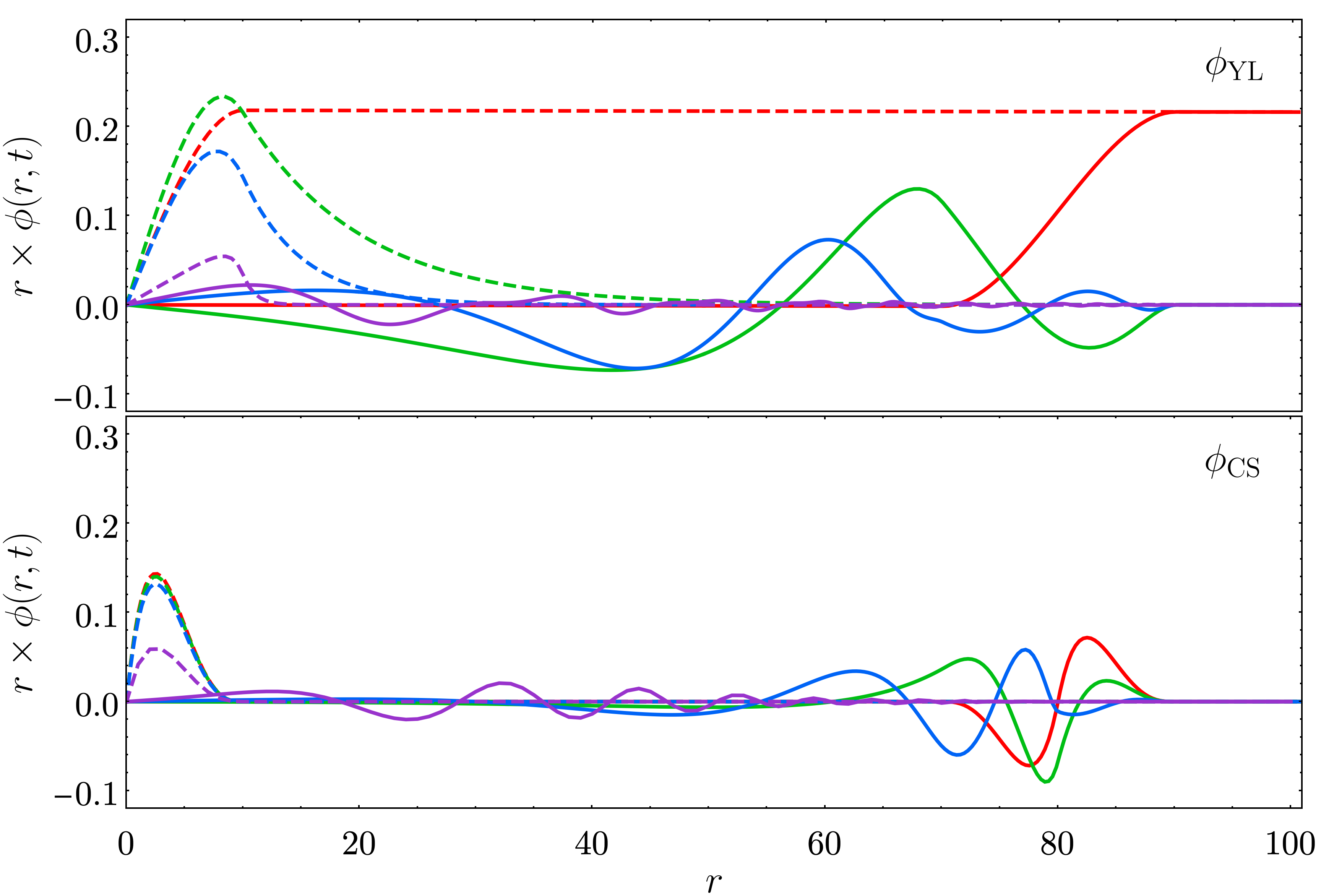}}
    \subfigure[\label{fig:field-evolution-2}]{\includegraphics[width=0.85\textwidth]{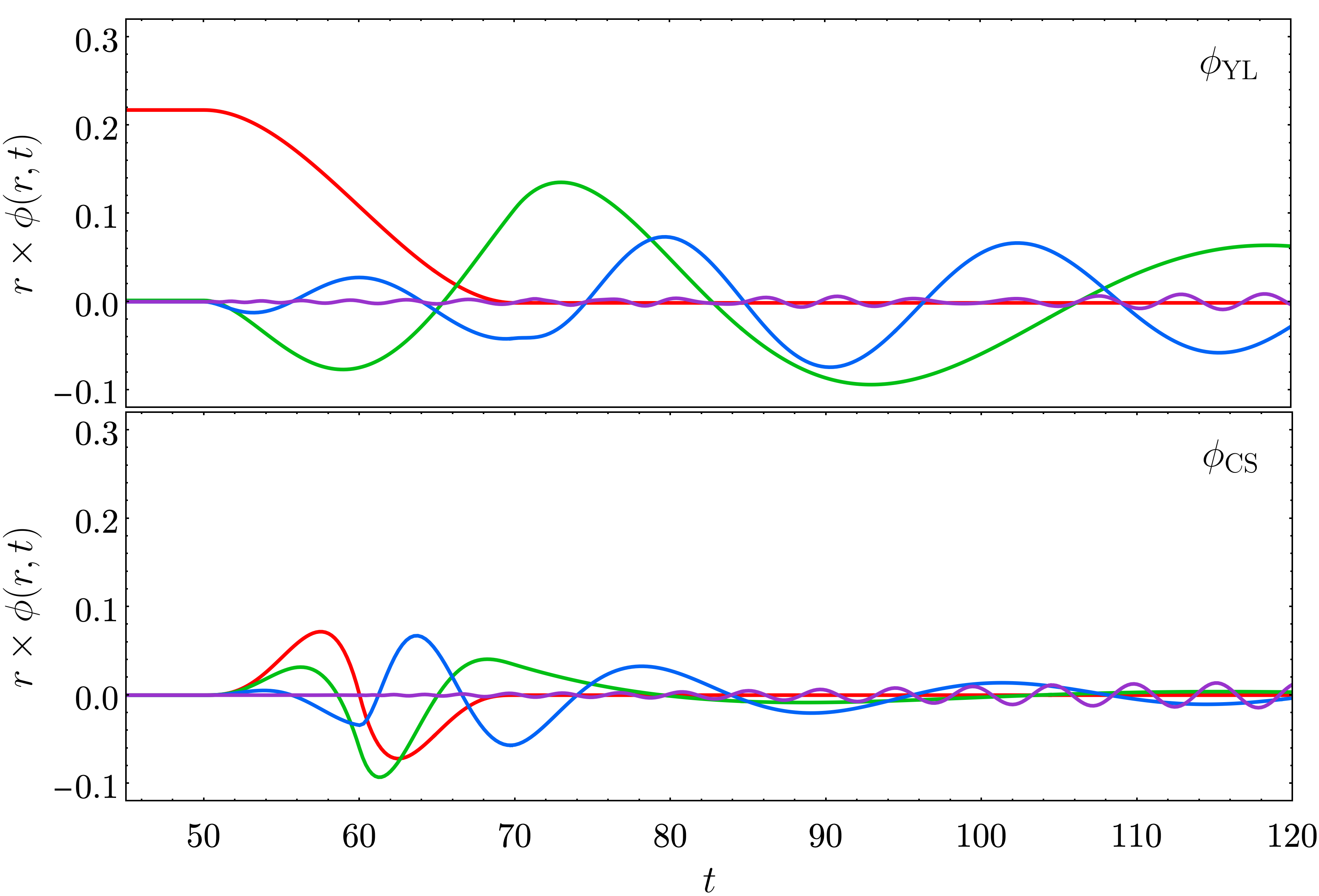}}
    \caption{Exemplary plot in arbitrary units of the time evolution of a propagating YL- and CS-field with $m_\phi=0$~(\myu{red}), $m_\phi=0.1$~(\myd{green}), $m_\phi=0.2$~(\myt{blue}) and $m_\phi=1$~(\myf{purple}). The radius has been fixed to $R=10$ and the energy to $E=0.1$. Note that we have multiplied the field value by $r$ for better visibility. \textbf{Top:} $t_1=0$~(dashed) and $t_2=80$~(solid). \textbf{Bottom: $r=60$.}}
    \label{fig-field-evolution}
\end{figure}

The time evolution of both YL- and CS- fields can be seen in Fig.~\ref{fig-field-evolution} for masses $m_\phi=\{0,\,0.1,\,0.2,\,1\}$ in arbitrary units. From this, we can see that the presence of a mass dilutes the signal in time. This typically limits the mass range in which Earth-bound experiments that are at a large distance from the source are sensitive.

\section{Experimental Sensitivities}
\label{sec:exp-sens}
In this section, we estimate the sensitivity of explicit experiments to transient signals as discussed in the previous section.
We consider the following exemplary interactions with the fermions and photons inside the detector
\begin{equation}
\label{eq:interactions}
    \mathcal{L}\subset -g_{S}\phi\,\ov{N}N+g_{S}\phi\,\ov{e} e -g_{P}\phi\,\ov{N}i\gamma_5N-\frac{g_{\gamma}}{4}\phi F^{\mu\nu} \widetilde{F}_{\mu\nu}\,.
\end{equation}
Here, $N=(p,n)^{T}$ denotes the nucleons (i.e. the protons $p$ and the neutrons $n$), $e$ the electrons and $F$ is the electromagnetic field strength tensor. The Lagrangian therefore specifies a $\sim (B-L)$-type scalar coupling to matter, a $\sim B$ coupling to the nucleon spins and a pseudo-scalar two-photon coupling.
While not exhaustive these couplings allow for signals in experiments employing quite different search strategies. For a wider range of possible couplings of scalar and vector particles see, e.g.~\cite{Beacham:2019nyx,Agrawal:2021dbo,Antypas:2022asj,Antel:2023hkf}.

In the next two subsections, we first give details of the sources we consider and the spectral density we will use to compare to the experimental observations. In the following ones, we will then go through the different types of coupling and a number of experiments to get an impression of the achievable sensitivities. 

\subsection{Sources and their Features}
\label{sec:sources}
As sources of our transients we consider four benchmarks for this work: the famous supernova SN1987A, the red giant supernova-candidate Betelgeuse, a distant neutron star merger (or a merger with a black hole) modelled on the LIGO event GW170817~\cite{KAGRA:2023pio}~\footnote{While the gravitational wave signal is compatible with the merger of two neutron stars, it is not clear whether the merging produced a new more massive neutron star or a very light black hole. We will assume the latter case.}. As an extreme example, we take the closest discovered Neutron Star~(NS) to Earth (which we assume to collapse into a black hole by an unspecified mechanism but one possibility could be due to an extra accretion of dark matter, see, e.g.~\cite{Garani:2018kkd}).
The relevant main properties of these objects are shown in Tab.~\ref{tab:sources}.
\begin{table}[tbh]
    \centering
    \begin{tabular}{cc|ccc}
    \toprule
     Name&  Type  & $M$~[$M_\odot$] & $r~[\text{ly}]$ & $R$\\
    \midrule
     SN1987A & Supernova & $20$& $170\times 10^3$ & $35\,R_\odot\approx 2.4\times 10^7~\text{Km}$\\
    Betelgeuse & Red Giant & 18 &$500$& $800\,R_\odot$\\
     LIGO event & Neutron stars merger & $(1.5$, $1.3)$&  $130\times 10^6$ & $10~\text{Km}$\\
     (GW170817)&&&&\\
         RX J1856.5-3754 & Neutron Star  & 0.9& $400$ & $10~\text{Km}$\\
    \bottomrule
    \end{tabular}
    \caption{Mass~($M$), distance~($r$) and radius~($R$) for our  four benchmarks events used through this work~\cite{Ho:2006uk,Smith:2006es, Joyce_2020, KAGRA:2023pio}. In reality, the relevant parameters feature sizable errors, but for our purposes, we only take them as order of magnitude benchmark values.}
    \label{tab:sources}
\end{table}

Before going into experimental details let us estimate the characteristic features of the expected signals.
The typical frequency of the signal is related to the size of the sources as 
\begin{equation}
    f_{\rm typ}\sim \frac{\omega_{typ}}{2\pi}\sim \frac{1}{2\pi R_{source}}.
\end{equation}
Evaluating this for the initial objects (star, supernova and neutron star) appearing in Tab.~\ref{tab:sources} we find,
\begin{align}
    &f_\text{Betelgeuse}\approx 9\times 10^{-5}\text{Hz}\,, &&f_\text{SN1987A}\approx2\times 10^{-3}\text{Hz}\,,
    &&f_\text{NS}\approx 5\times 10^{3}\text{Hz}\,.
\end{align}

As we have seen from Fig.~\ref{fig-field-evolution}, we also need a sufficiently small mass in order for the signal to not spread out too much in time. Roughly speaking this requires that the velocity of the particles emitted with different frequencies does not differ too much such that the spatial distance $r$ between them is much larger than the size of the initial signal, i.e. the size of the source,
\begin{equation}
\Delta v_{typ} r\lesssim R.
\end{equation}
As the width in frequency is of the same order as the typical frequency, this requires,
\begin{equation}
\label{eq:mass-distance-condition}
    \frac{m_{\phi}}{\omega_{typ}}r\sim m_{\phi}R r\lesssim R 
\end{equation}
and hence,
\begin{equation}
m_{\phi}\lesssim \frac{1}{r}.
\end{equation}

Depending on the distance of the source this restricts the mass reach of the different types of transients to be,
\begin{align}
     &m_\phi^\text{SN1987A}\lesssim \frac{1}{r_\text{SN1987A}}\approx1\times 10^{-28}\eV\,, &m_\phi^\text{Betelgeuse}\lesssim \frac{1}{r_\text{Betelgeuse}}\approx 4\times 10^{-26}\eV\,,\\
     &m_\phi^\text{NS}\lesssim \frac{1}{r_\text{NS}}\approx 5\times 10^{-26}\eV\,, & m_\phi^\text{Merger}\lesssim \frac{1}{r_\text{Merger}}\approx 2\times 10^{-31}\eV\,.
\end{align}

\subsection{Spectral Density of the Signals}
After the preliminary estimates above, let us characterize the signals a bit more precisely in terms of the power spectrum. We can then use this to compare directly to the experimental noise curves.

We start with the
Fourier Transform~(FT) of a signal function $h(t)$, 
    \begin{equation}
        \Tilde{h}(\omega)\equiv \int\limits_{-\infty}^\infty \diff{t}\,h(t)e^{-i\omega t}\,.
    \end{equation}
The Amplitude and Power Spectral Density~(PSD and ASD, respectively) are given by
\begin{align}
    &\text{ASD}[h](f)\equiv \sqrt{f}|\Tilde{h}(f)|\,, & \text{PSD}[h](f)\equiv |\text{ASD}[h](f)|^2\,,
\end{align}
where the frequency is simply given by $f\equiv \omega/(2\pi)$.
Based on matched filtering (cf., e.g.~\cite{optimal}, but also~\cite{Daykin:2021lzn} for a more advanced discussion), the sensitivity of an experiment to the signal can be estimated by comparing the signal-PSD with the noise power spectral density, $S(f)$.
Finally, the signal-to-noise-ratio~(SNR) is defined as~\cite{PhysRevD.85.122006}
    \begin{equation}
    \label{eq:SNR-definition}
        \text{SNR}^2 \equiv 4\int\limits_{0}^\infty\diff{f}\dfrac{|\tilde{h}(f)|^2}{S(f)}=4\int\limits_{0}^\infty\diff{\log{f}}\, \dfrac{\text{PSD}[h](f)}{S(f)}\,.
    \end{equation}

\begin{figure}
    \centering
    \includegraphics[width=\textwidth]{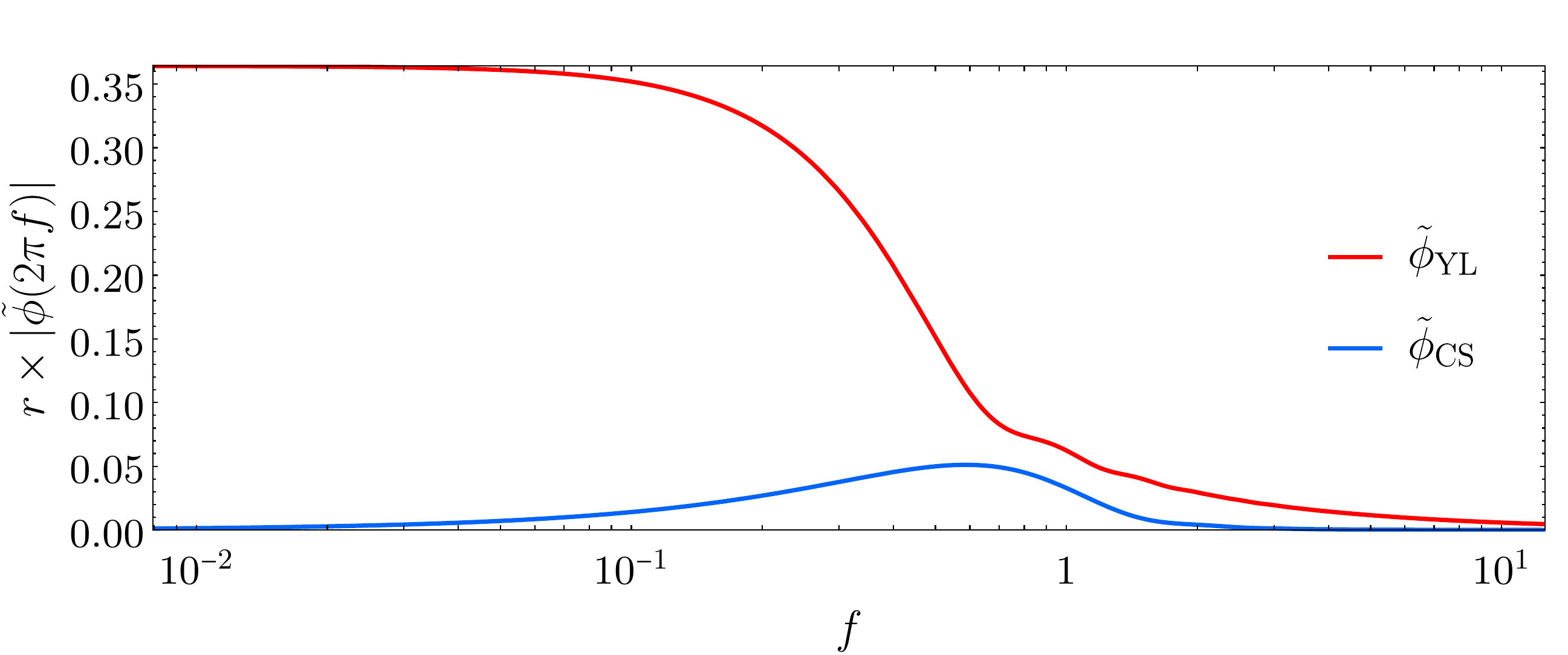}
    \caption{Illustrative plot of the FT of a YL- and a CS-field as a function of the frequency with $R=1$ and $|E|=1$ in arbitrary units. Note that we plot as a function of $f=\omega/(2\pi)$.}
    \label{fig:FT-signals}
\end{figure}

For later convenience let us examine the FT of the field in the two cases of interest. We consider only the massless case and limit our sensitivity estimates to the range of masses where this is a good approximation. 
Different observables which are induced by spatial or time derivatives of the field can then be computed straightforwardly by employing FT properties. At leading order in the $R/r$ expansion, the FTs for our two benchmark shapes read,
\begin{align}
    &\Tilde{\phi}_\text{YL}(\omega)\approx\sqrt{\frac{5}{3 \pi }} e^{-i \omega  (r+R)}\times  \frac{ e^{2 i \omega R } \left(3-i \omega R  \left(2 R^2 \omega ^2+3\right)\right)-3 i \omega R -3}{4 r R^3 \omega ^4}\,,\\
    &\Tilde{\phi}_\text{CS}(\omega)\approx i \sqrt{\frac{210}{\pi }} e^{-i \omega r }\times \frac{ R^2 \omega ^2+\omega R  \sin (\omega R )+4 \cos (\omega R )-4}{r R^4 \omega ^5}\,.
\end{align}
In the limit of $\omega R\ll1$ and $\omega R\gg 1$ they scale as
\begin{align}
    &\dfrac{\Tilde{\phi}_\text{YL}(\omega)}{\sqrt{|E|R}}\approx \sqrt{\dfrac{5}{12\pi}}\dfrac{R }{r}e^{-i\omega r}\times\begin{cases}
        1& \omega R\ll1\,,\\
        -i\dfrac{e^{i \omega R}}{\omega R}& \omega R \gg 1\,,
    \end{cases}\\
    &\dfrac{\Tilde{\phi}_\text{CS}(\omega)}{\sqrt{|E|R}}\approx i\sqrt{\dfrac{7}{30\pi}}\dfrac{R }{r} e^{-i\omega r}\times\begin{cases}
      \dfrac{\omega R}{12}  & \omega R\ll1\,,\\
      &\\
        \dfrac{30}{(\omega R)^3}& \omega R \gg 1\,.
    \end{cases}
\end{align}
The difference between YL and CS sources is evident from the above equations and an illustrative plot of their shapes is given in Fig.~\ref{fig:FT-signals}. The YL spectrum is suppressed for large frequencies $2\pi fR=\omega R\gtrsim 1$ but tends to be constant for small ones. On the contrary, the CS has a frequency spectrum peaked around $2\pi fR=\omega R\sim 5$ and is suppressed for both frequencies much larger and smaller than $R$. We note that the peak is around $2\pi fR=\omega R\sim 5$ instead of $2\pi fR=\omega R\sim 1$ because the shape for our CS configuration, Eq.~\eqref{eq:CS-benchmark}, actually has most of the field concentrated around the center. This skews the distribution of frequency to slightly higher values. This is effectively a choice in what we call the size of the compact source. For consistency, we nevertheless account for this factor also in the typical mass reach when considering the CS by modifying Eq.~\eqref{eq:mass-distance-condition}
\begin{equation}
    m_{\phi,\text{CS}}\lesssim \frac{5}{r}\,.
\end{equation}

We are now ready to consider the different couplings and experiments that could be sensitive to the discussed transient signals. In the following subsections we will consider each of the couplings in Eq.~\eqref{eq:interactions} and look at suitable experiments to probe them.

\subsection{Scalar Coupling to Matter}


For the scalar couplings in Eq.~\eqref{eq:interactions} gradients in the field cause forces on the matter that it is coupled to (cf., e.g.~\cite{Graham:2015ifn}),
\begin{equation}
    \vec{F}\sim -\vec{\nabla}\phi.
\end{equation}
Therefore, time- and space-dependent field configurations cause a corresponding force on matter objects.
Hence, suitable experiments are the ones that are sensitive to time-dependent forces.

\subsubsection{Interferometer: LISA Pathfinder}
A suitable type of experiments are ones sensitive to the drifting of probe masses such as LISA Pathfinder~\cite{LISAPathfinder:2017khw}.

Denoting by $M$ the probe mass and by $q$ its total charge under the new scalar field, the evolution of its position in time is dictated by the second Newton's law
\begin{equation}
    a(t)=g_S\left(\frac{q}{M}\right)\left(-\partial_r \phi(r,t)\right)\,.
\end{equation}

We adopt the ASD sensitivity of Ref.~\cite{LISAPathfinder:2018sdh}.\footnote{We note that for setting an exclusion limit one may need to be very careful with the noise suppression measures taken in the data analysis~\cite{private}.} 
The direction of the signal is unknown and the noises for different directions are slightly different but do not change the order of magnitude of the result. As a concrete example, we employ the noise data derived for the x-axis; a different choice does not change the results qualitatively. As already specified in Eq.~\eqref{eq:interactions} we consider a $B-L$-type coupling. This coupling structure is analogue to the DM benchmark studied in Ref.~\cite{Frerick:2023xnf} with the same experiment. In this case, the effective charge-to-mass ratio for the force difference between the compared test objects is given by $\Delta(q/M)=0.018$~\cite{Frerick:2023xnf}.

\begin{figure}[t]
    \centering
    \subfigure[{}\label{fig:LISA-ASD-KG}]{\includegraphics[width=0.49\textwidth]{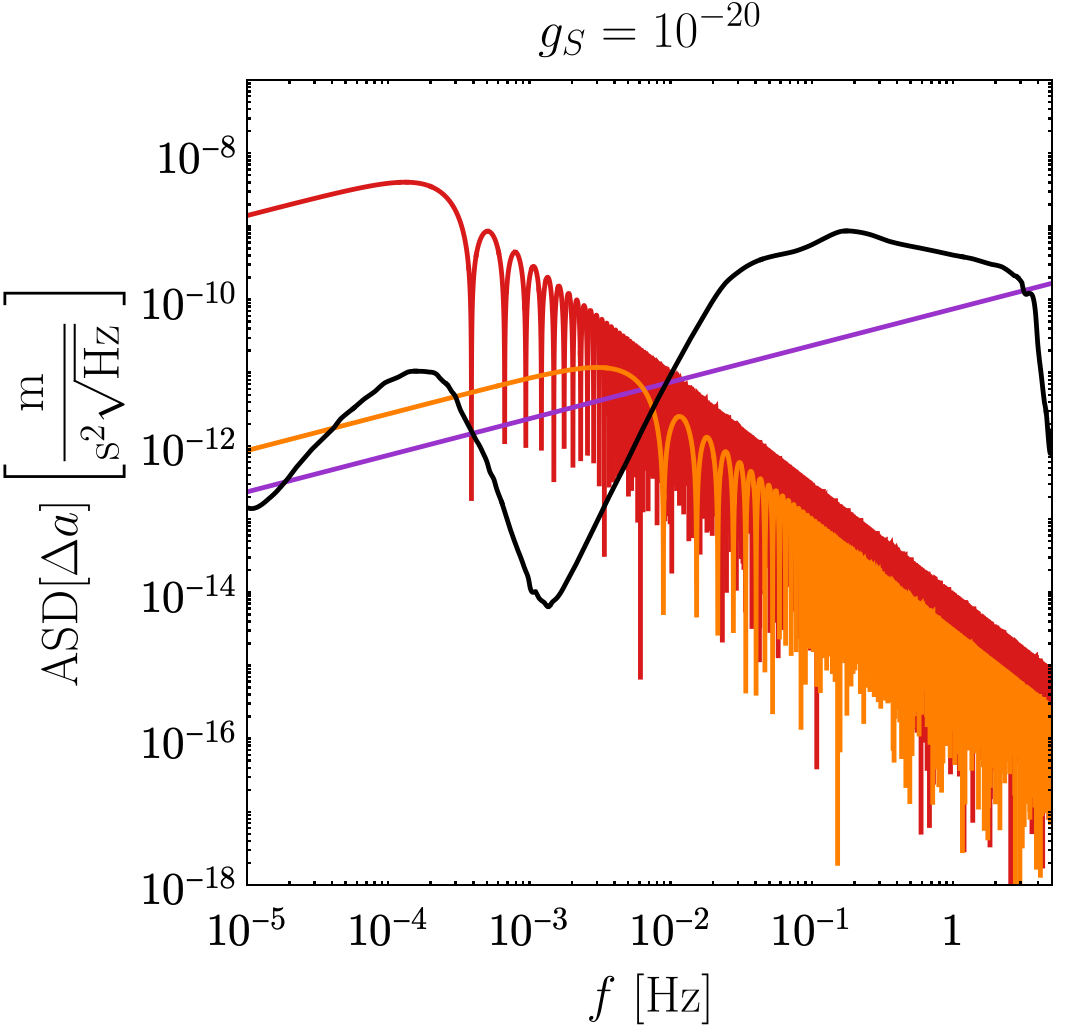}}\, 
    \subfigure[{}\label{fig:LISA-ASD-CS}]{\includegraphics[width=0.49\textwidth]{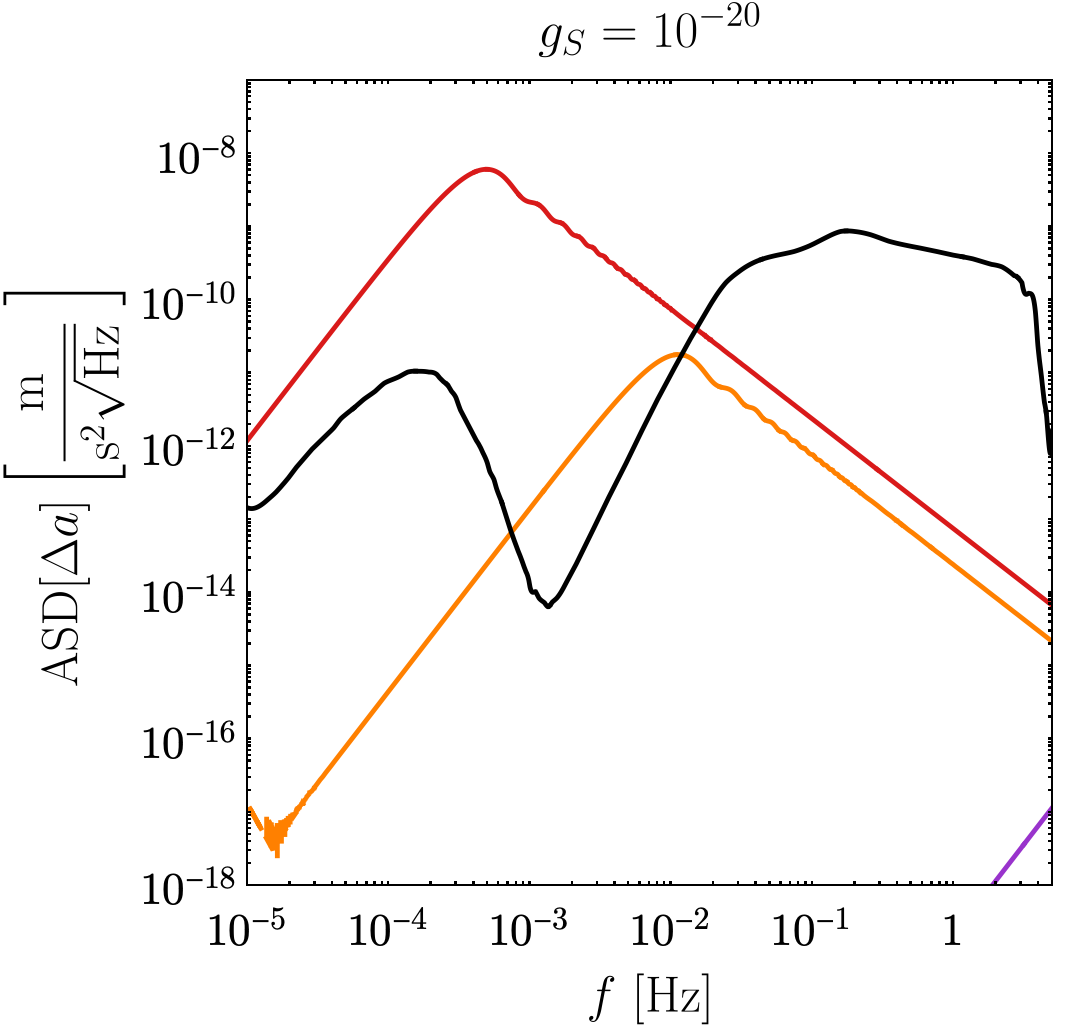}}
    \caption{ASD of the relative acceleration induced by the signals (red giant~(\redG{red}), supernova~(\orangeG{orange}) and neutron star~(\purpleG{purple})) compared to the sensibility of LISA Pathfinder~\cite{LISAPathfinder:2018sdh} (black) for a YL-~(left) and CS-source~(right). The coupling is fixed to $g_S=10^{-20}$ and the total field energy is set to $E= 0.1\,M_\odot$.  }
    \label{fig:LISA-ASD}
\end{figure}

The Fourier transforms of the gradient of the field, $\widetilde{F}$, at LO in the large-$r$ limit read
\begin{align}
    &\widetilde{F}_\text{YL}(\omega)\approx\sqrt{\frac{15}{\pi }} e^{-i r \omega }  \frac{\omega R  \cos (\omega R)-\sin ( \omega R)}{2 r  (\omega R)^3}\,,\\
    &\widetilde{F}_\text{CS}(\omega)\approx\sqrt{\frac{210}{\pi }} e^{-i r \omega }\frac{ (\omega R)^2+ \omega R \sin ( \omega R)+4 \cos (\omega R)-4}{r (\omega R)^4}\,.
\end{align}
Their expansion in the low- and high-frequency limit is given by
\begin{align}
    &\widetilde{F}_\text{YL}(\omega)\approx-\frac{\sqrt{ER}}{r}\times\sqrt{\frac{5}{12\pi}}\begin{cases}
        1 & \omega R\ll 1\,,\\
        -\frac{3\cos(\omega R)}{(\omega R)^2} & \omega R\gg 1\,,
    \end{cases}\\
    &\widetilde{F}_\text{CS}(\omega)\approx-\frac{\sqrt{ER}}{r}\times\sqrt{\frac{7}{30\pi}}\begin{cases}
        (\omega R)^2/12 & \omega R\ll 1\,,\\
        \frac{30}{(\omega R)^2} & \omega R\gg 1\,.
    \end{cases}
\end{align}
In SI units, the relative acceleration ASD is then found to be
\begin{equation}
    \text{ASD}[\Delta a]=g_S\left(\Delta\frac{q}{M}\right)(\hbar)\times\sqrt{f}\left|\widetilde{F}\left(\frac{2\pi f}{c}\right)\right|\,.
\end{equation}
Its comparison for the two types of signals with the background can be seen in Fig.~\ref{fig:LISA-ASD}.
The signal starts decreasing for frequencies larger than the characteristic frequency of the source $\sim R^{-1}$, in agreement with the analytic estimations.
\begin{table}[H]
    \centering
    \begin{tabular}{c|cccccc}
    \toprule
    Experiment & $g_{S,\text{YL}}^\text{Red Giant}$ & $g_{S,\text{CS}}^\text{Red Giant}$ & 
$g_{S,\text{YL}}^\text{SN1987A}$ & 
$g_{S,\text{CS}}^\text{SN1987A}$ & 
$g_{S,\text{YL}}^\text{NS}$ & 
$g_{S,\text{CS}}^\text{NS}$ 
\\
    \midrule
     LISA Pathfinder    & $\bm{3\times 10^{-25}}$ & $\bm{3\times 10^{-26}}$ & $\bm{5\times 10^{-24}}$ & $1\times 10^{-22}$ & $2\times 10^{-23}$ & $1.6\times 10^{-15}$   \\
     Dual Oscillator    & $6\times 10^{-24}$ & $6\times 10^{-25}$ & $2\times 10^{-23}$ & $\bm{2\times 10^{-24}}$ & $\bm{7\times 10^{-26}}$ & $\bm{5\times 10^{-19}}$  \\
    \bottomrule
    \end{tabular}
    \caption{Sensitivities on $|g_S|$ from the LISA Pathfinder (cf. Eq.s~\eqref{eq:B-LPF-1}, \eqref{eq:B-LPF-2} and \eqref{eq:B-LPF-3}) and the dual torsion pendulum (cf. Eqs.~\eqref{eq:B-TP-1}, \eqref{eq:B-TP-2} and \eqref{eq:B-TP-3}) experiments. The strongest sensitivities for each case are highlighted in bold font. The sensitivities for the mergers can be obtained by rescaling the ones of the NS-case. The field energy release was fixed to $E=0.1\,M_\odot$. The bold values are those shown in Fig.~\ref{fig:coupling-bounds}.}
    \label{tab:scalar-bounds}
\end{table}
\begin{figure}[]
    \centering

    \includegraphics[width=\textwidth]{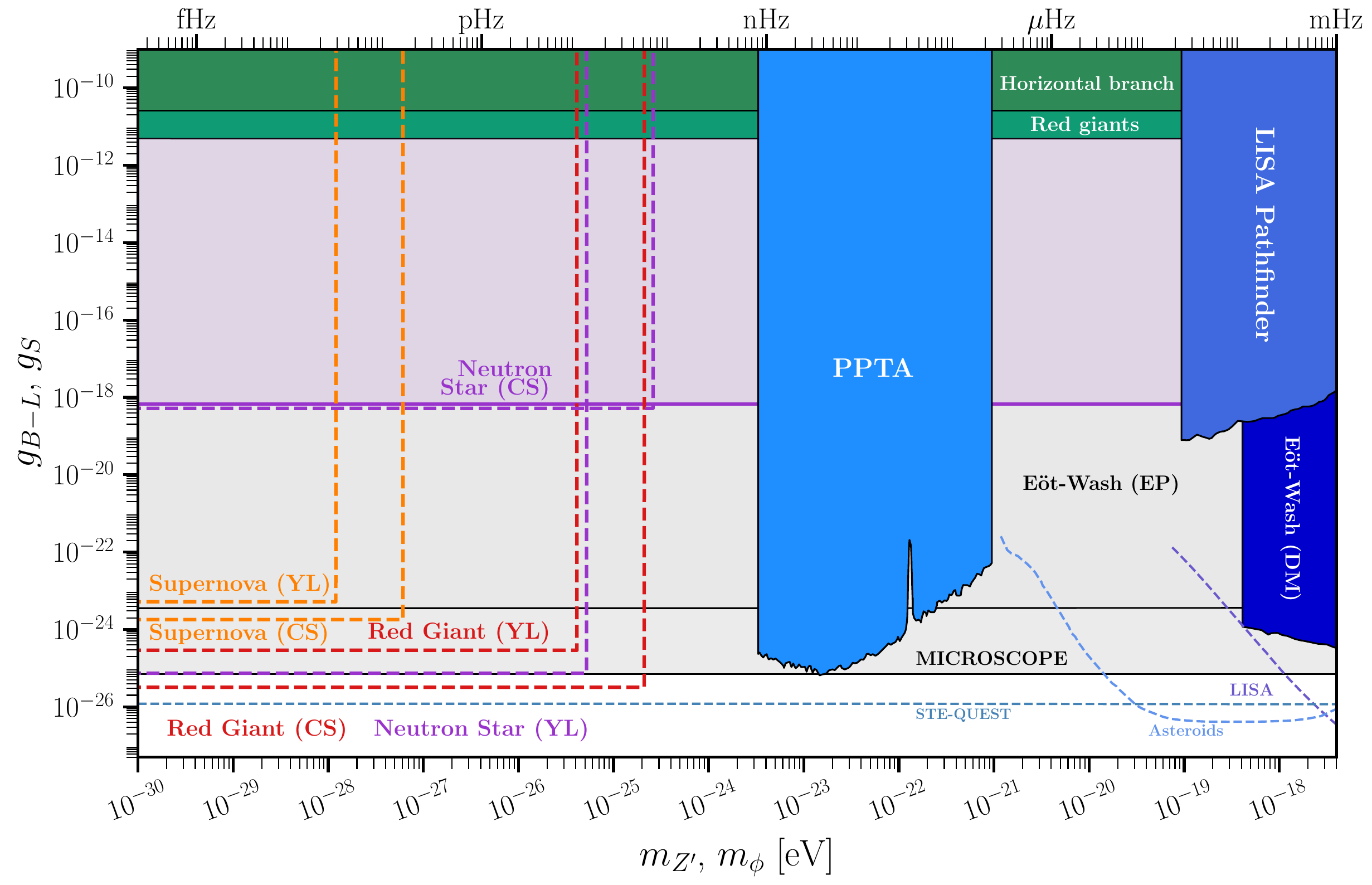}
    \caption{Strongest sensitivities from Table~\ref{tab:scalar-bounds} to $|g_S|$ for a YL- and CS-sources adopting $E=10^{-1}\,M_\odot$ for the released field energy. We compare them to existing ones on $(B-L)$ coupling from astrophysical sources~(green), laboratory tests of the equivalence principle~(grey) and current~(solid) and projected~(dashed) bounds from DM searches~(blue). The solid purple line shows the constraint arising for a Yukawa coupling where the field energy in the Yukawa field at the source exceeds $E\geq 0.5 M$ for the neutron star; similar limits for the other objects are weaker. Plot adapted from Refs.~\cite{OHare:2020wah,ohare}. The main limits and projections shown are from~\cite{Li:2023vpv,Wagner:2012ui, MICROSCOPE:2022doy, Amaral:2024tjg,PPTA:2021uzb,Shaw:2021gnp, Miller:2023kkd,STE-QUEST:2022eww,Fedderke:2022ptm}.}
    \label{fig:coupling-bounds}
\end{figure}
    
Integrating over the full available range of frequencies and requiring $\text{SNR} \leq 1$\footnote{$\text{SNR}=1$ is a rather low value. But we stress again that our considerations are only order of magnitude.} we obtain
\begin{align}
    &\label{eq:B-LPF-1}g^\text{Red Giant}_\text{S}\lesssim \begin{Bmatrix}
       3\times 10^{-25}\,, & \text{YL}\\
       3\times 10^{-26}\,, & \text{CS}
    \end{Bmatrix}\times\left(\sqrt{\dfrac{10^{-1}\,M_\odot}{E}}\dfrac{r}{500\,[\text{ly}]}\right)\,,\\
    &\label{eq:B-LPF-2}g^\text{SN1987A}_\text{S}\lesssim \begin{Bmatrix}
       5\times 10^{-24}\,, & \text{YL}\\
       1\times 10^{-22}\,, & \text{CS}
    \end{Bmatrix}\times\left(\sqrt{\dfrac{10^{-1}\,M_\odot}{E}}\dfrac{r}{1.7\times 10^{5}\,[\text{ly}]}\right)\,,\\
    &\label{eq:B-LPF-3}g^\text{NS}_\text{S}\lesssim \begin{Bmatrix}
       2\times 10^{-23}\,, & \text{YL}\\
       2\times 10^{-15}\,, & \text{CS}
    \end{Bmatrix}\times\left(\sqrt{\dfrac{10^{-1}\,M_\odot}{E}}\dfrac{r}{400\,[\text{ly}]}\right)\,,\\
    &\,\,\,\quad\nonumber\lesssim\begin{Bmatrix}
       6\times 10^{-18}\,, & \text{YL}\\
       5\times 10^{-10}\,, & \text{CS}
    \end{Bmatrix}\times\left(\sqrt{\dfrac{10^{-1}\,M_\odot}{E}}\dfrac{r}{1.3\times 10^{8}\,[\text{ly}]}\right)\,.
\end{align}

As we can see, the limits for the YL and the CS are in the same ballpark for the Red Giant and the supernova case, but they differ considerably for the neutron star. The reason is that the former two have a typical frequency that matches reasonably well to the sensitive frequencies of the experiment. In contrast in the neutron star case, the typical frequency is much higher and we therefore depend on the low frequency part of the spectrum which is strongly suppressed for the CS. This is because the field completely vanishes outside the source in this case. In the YL the field also features a $\sim 1/r$ component that extends to larger distances from the source.\footnote{We also note that the relevant distances in the present case are of the order of a few to several hundred light seconds. This is of the order of an astronomical unit and therefore not extremely large on the scale of a star system.}

\subsubsection{Torsion Pendulum}
        A torsion pendulum subject to a time-dependent torque, $\tau(t)$, obeys the following equation of motion
        \begin{equation}
        \label{eq:torque-eq}
            \ddot{\theta}+\left(\dfrac{\omega_0}{Q}\right)\dot{\theta}+\omega_0^2\theta = \dfrac{\tau(t)}{I}\,,
        \end{equation}
        where $Q$ is the quality factor, $\omega_0$ is the proper pulsation of the pendulum and $I$ is its rotational inertia.
         We consider the apparatus of Ref.~\cite{Shaw:2021gnp}, whose pendulum was suspended by a fused-silica fiber with resonant frequency $f_0 \equiv (2\pi)^{-1}\omega_0 = 1.934$~mHz, $I = 3.78 \times 10^{-5}~\text{Kg m}^2$ and $Q = 4.6\times 10^5$. The experiment has searched for a DM signal, which induces a cosine-like torque of the type
         \begin{equation}
             \tau_\text{DM}(t)\propto \cos{(m_\text{DM}t)}
         \end{equation}
Data have been collected throughout $T=114$~days.
The damping of the signal is expected to be relevant in a time scale of 
\begin{equation}
t_\text{damping}\sim\frac{Q}{\omega_0}\sim 10^7~\text{s}\sim 400~\text{days}\,,
\end{equation}
which is much larger than the measurement time. The signal detected in the experiment is expected to grow over time towards the steady periodic state which, however, is not reached in the measurement time. To estimate the sensitivity, we, therefore, simulate the signal of Ref.~\cite{Shaw:2021gnp} by assuming a DM-like signal and solving Eq.~\eqref{eq:torque-eq}. The Fourier Transform is then computed truncating the integral at $T=114$~days
\begin{equation}
    \widetilde\theta_\text{DM}(\omega)=\int\limits_0^T\diff{t}\, e^{-i\omega t}\theta_\text{DM}(t)\,.
\end{equation}
We compare its power spectrum with the expected one from the transient signal. The Fourier Transform of the angle can be read from Eq.~\eqref{eq:torque-eq}
\begin{equation}
    \widetilde\theta(\omega)=\dfrac{1}{I}\dfrac{\widetilde\tau(\omega)}{\omega_0^2-\omega^2+i\left(\dfrac{\omega\omega_0}{Q}\right)}\,.
\end{equation}

\begin{figure}[]
    \centering
    \subfigure[{}\label{fig:torque-ASD-KG}]{\includegraphics[width=0.49\textwidth]{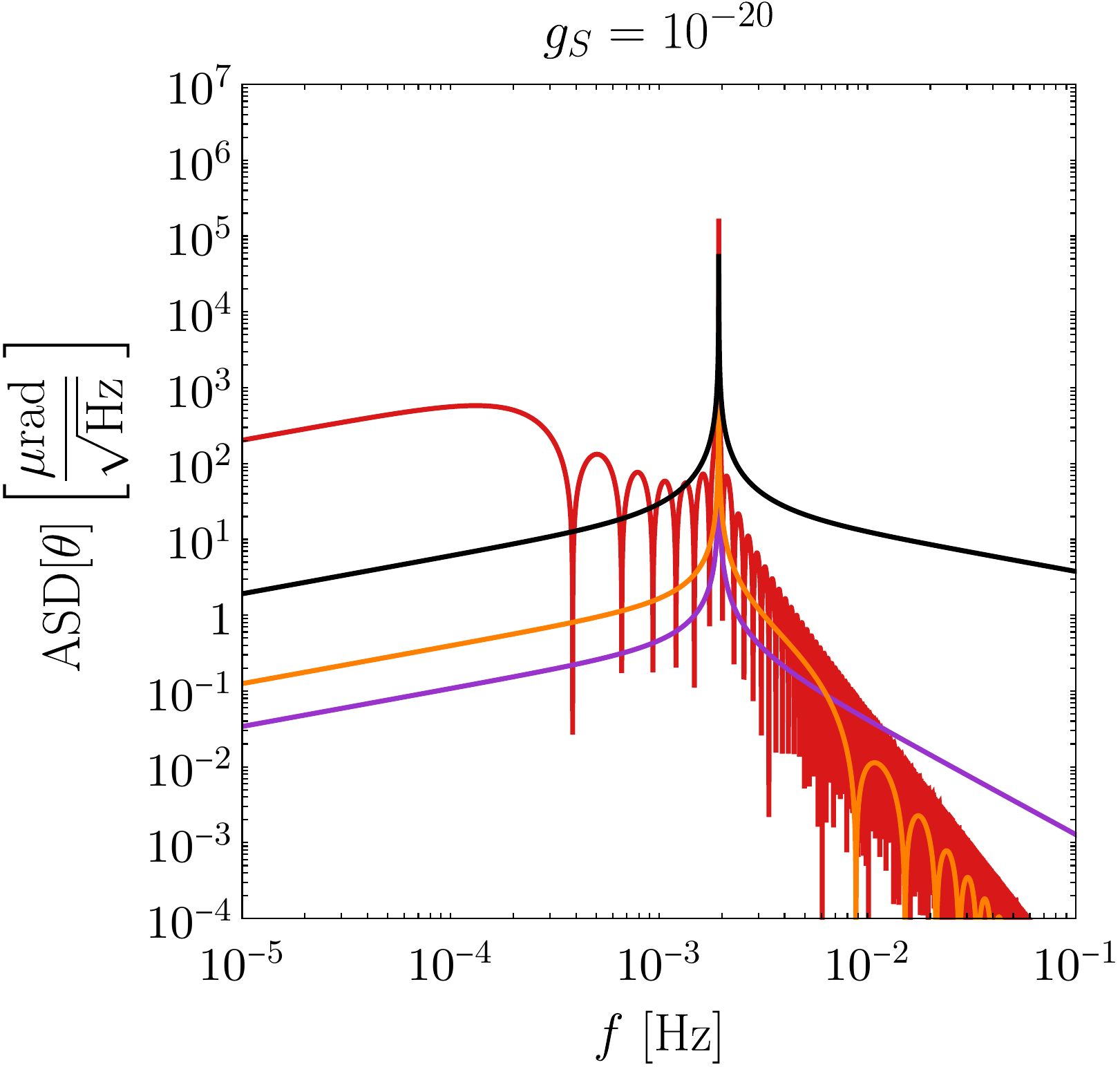}}
    \subfigure[{}\label{fig:torque-ASD-CS}]{\includegraphics[width=0.49\textwidth]{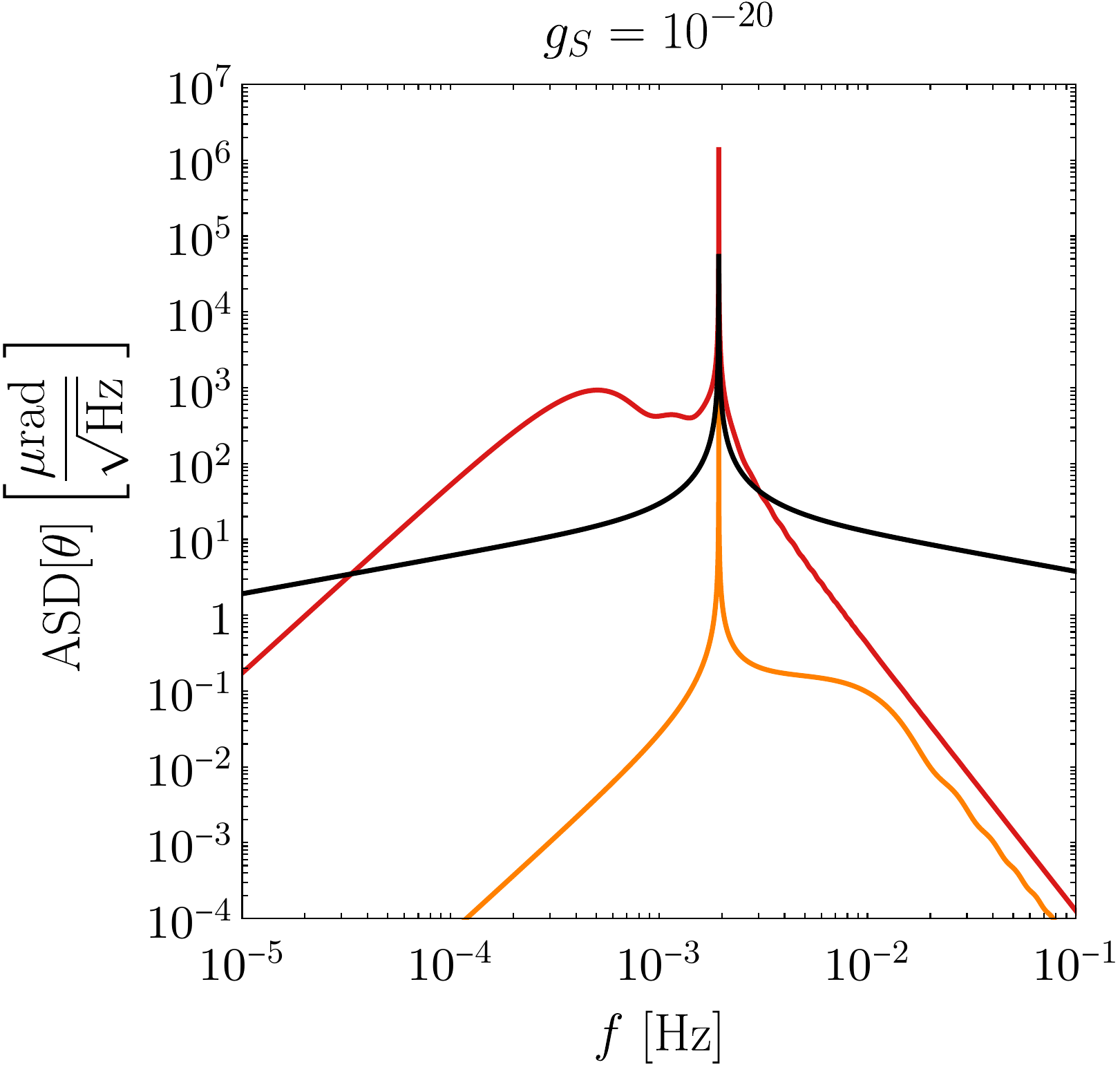}}
    \caption{ASD of the angle induced by the signals (red giant~(\redG{red}), supernova~(\orangeG{orange}) and neutron star~(\purpleG{purple})) compared to the predicted angle of Ref.~\cite{Shaw:2021gnp}~(black) for a YL-~(left) and CS-source~(right). The DM frequency was taken to be on resonance, $\omega_\text{DM}=\omega_0$ and the coupling to be $g_\text{B-L}=10^{-25}$. The coupling of the signal was fixed to $g_S=10^{-20}$ and $E=10^{-1} M_\odot$.}
    \label{fig:torque-ASD}
\end{figure}
The result can be seen both for a YL- and CS-signal in Fig.~\ref{fig:torque-ASD}. Assuming the maximum value of the already constrained coupling $g_\text{B-L}=10^{-25}$, the signals are of comparable size at a coupling value of $g_S=10^{-20}$, which suggests sensitivities of this order of magnitude. This is not competitive with the ones in the previous section.

\subsubsection{Torsion Pendulum Dual Oscillator}
\begin{figure}[]
    \centering
    \subfigure[{}\label{fig:dualtorque-ASD-KG}]{\includegraphics[width=0.49\textwidth]{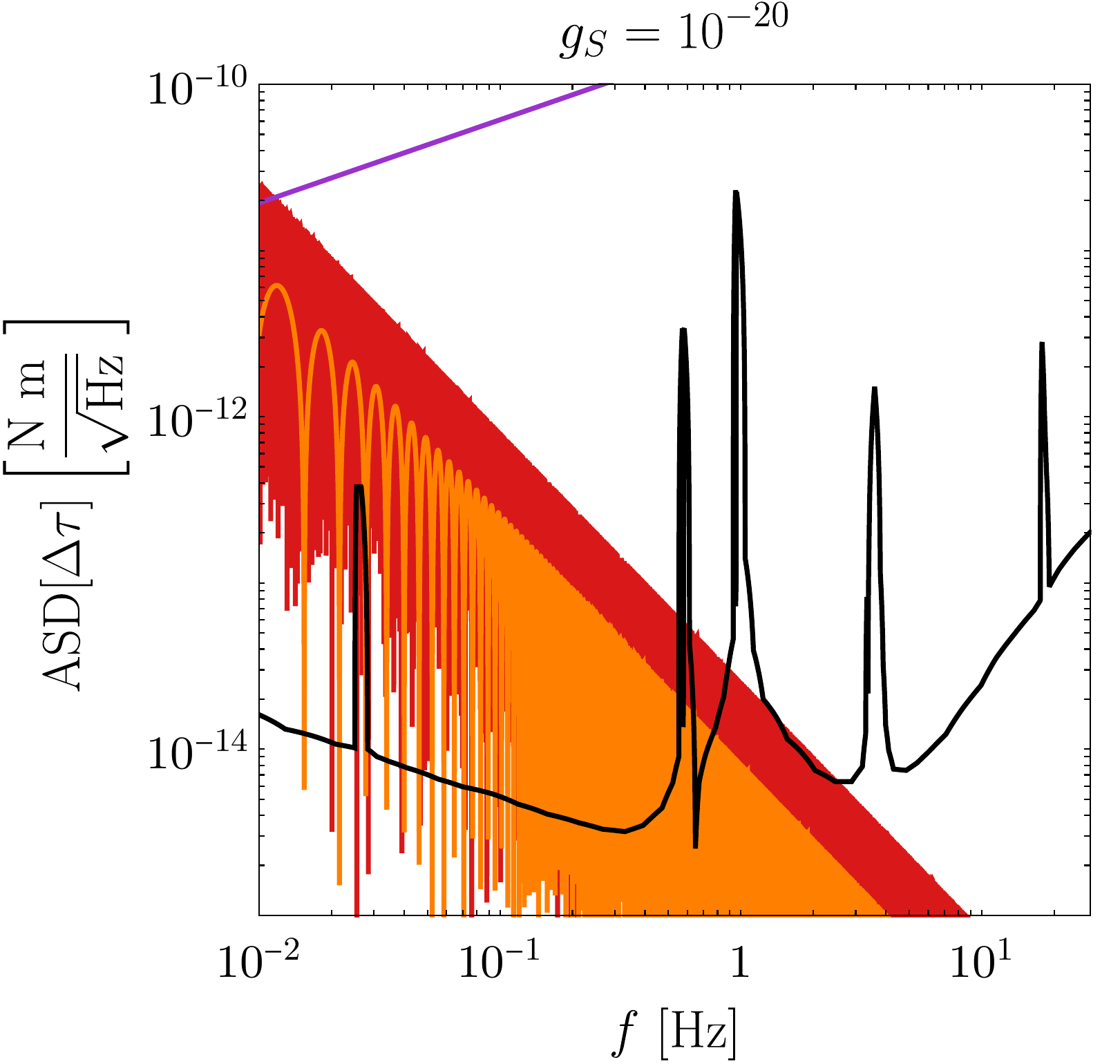}}
    \subfigure[{}\label{fig:dualtorque-ASD-CS}]{\includegraphics[width=0.49\textwidth]{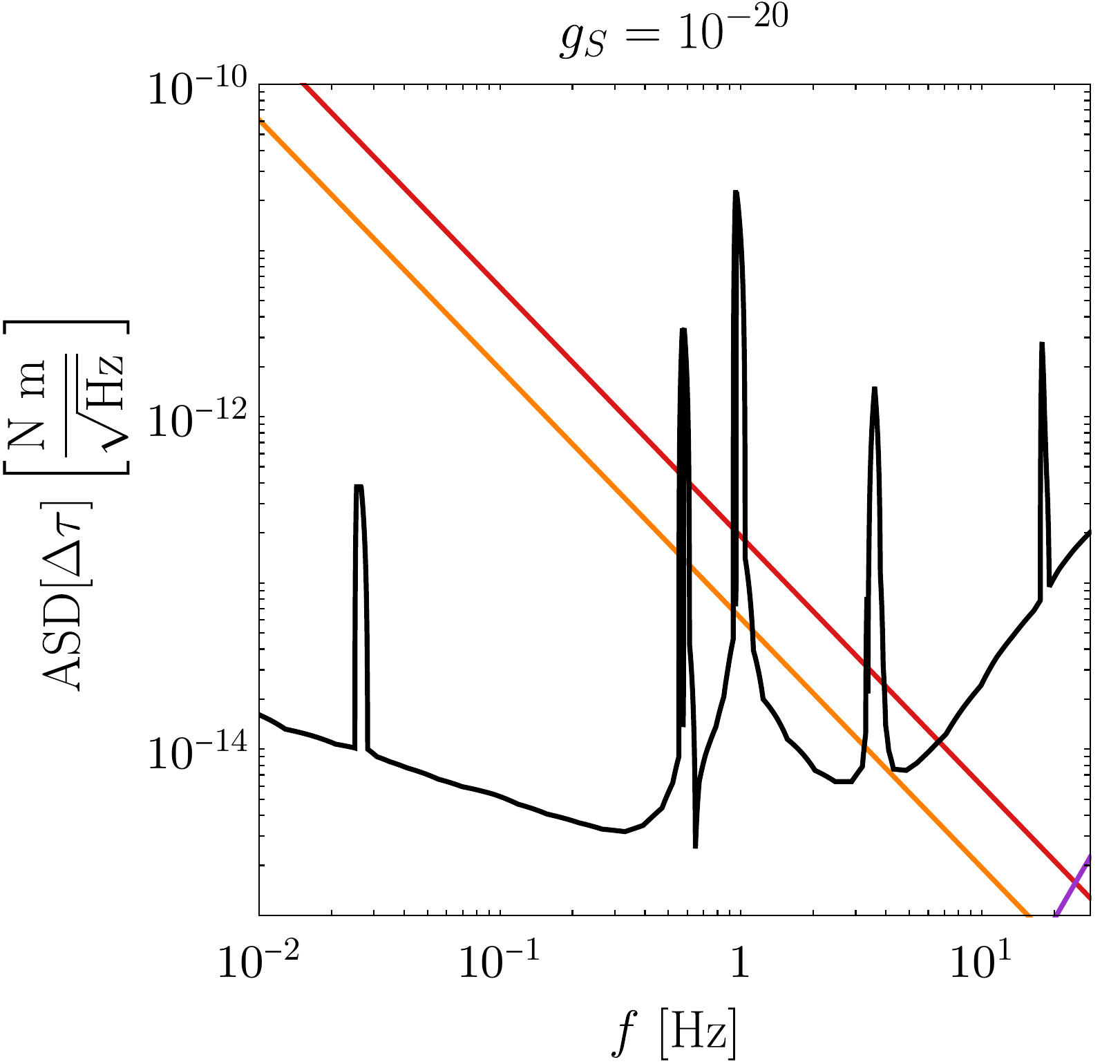}}
    \caption{ASD of the angle induced by the signals (red giant~(\redG{red}), supernova~(\orangeG{orange}) and neutron star~(\purpleG{purple})) compared to the noise ASD of TorPeDO from Ref.~\cite{Sun:2024qis}~(black) for a YL-~(left) and CS-field~(right). The coupling of the signal was fixed to $g_S=10^{-20}$ and $E=10^{-1} M_\odot$. }
    \label{fig:torque-dual-ASD}
\end{figure}

Let us also consider the projected sensitivity of a torsion pendulum dual oscillator (TorPeDO) as proposed in Ref.~\cite{Sun:2024qis} which estimates the sensitivity of a modified TorPeDO sensor to an ultralight dark matter field coupled to $B-L$ with masses $m_\text{DM}\in[10^{-17},10^{-13}]$~eV. The comparison of the differential torque induced by our transient signals and the background noise can be seen in Fig.~\ref{fig:torque-dual-ASD}.

Finally, the sensitivities from SNR read
\begin{align}
    &\label{eq:B-TP-1}g^\text{Red Giant}_\text{S}\lesssim \begin{Bmatrix}
       6\times 10^{-24}\,, & \text{YL}\\
       6\times 10^{-25}\,, & \text{CS}
    \end{Bmatrix}\times\left(\sqrt{\dfrac{10^{-1}\,M_\odot}{E}}\dfrac{r}{500\,[\text{ly}]}\right)\,,\\
    &\label{eq:B-TP-2}g^\text{SN1987A}_\text{S}\lesssim \begin{Bmatrix}
       2\times 10^{-23}\,, & \text{YL}\\
       2\times 10^{-24}\,, & \text{CS}
    \end{Bmatrix}\times\left(\sqrt{\dfrac{10^{-1}\,M_\odot}{E}}\dfrac{r}{1.7\times 10^{5}\,[\text{ly}]}\right)\,,\\
    &\label{eq:B-TP-3}g^\text{NS}_\text{S}\lesssim \begin{Bmatrix}
       7\times 10^{-26}\,, & \text{YL}\\
       5\times 10^{-19}\,, & \text{CS}
    \end{Bmatrix}\times\left(\sqrt{\dfrac{10^{-1}\,M_\odot}{E}}\dfrac{r}{400\,[\text{ly}]}\right)\,,\\
    &\,\,\,\quad\nonumber\lesssim\begin{Bmatrix}
       2\times 10^{-20}\,, & \text{YL}\\
       2\times 10^{-13}\,, & \text{CS}
    \end{Bmatrix}\times\left(\sqrt{\dfrac{10^{-1}\,M_\odot}{E}}\dfrac{r}{1.3\times 10^{8}\,[\text{ly}]}\right)\,.
\end{align}

The resulting sensitivities alongside the ones obtained in Eqs.~\eqref{eq:B-LPF-1}, \eqref{eq:B-LPF-2}, \eqref{eq:B-TP-2}, \eqref{eq:B-TP-3} are summarized in Table~\ref{tab:scalar-bounds} and shown in Fig.~\ref{fig:coupling-bounds} for $E=10^{-1}\,M_\odot$. The colour scheme encodes red giant~(\redG{red}), supernova~(\orangeG{orange}) and neutron star~(\purpleG{purple}). We compare such results with existing ones in the literature on a $(B-L)$ coupling from astrophysical sources~\cite{Li:2023vpv}~(green) and laboratory tests of the equivalence principle~\cite{Wagner:2012ui, MICROSCOPE:2022doy, Amaral:2024tjg}~(grey). For completeness, we also report current~\cite{PPTA:2021uzb,Shaw:2021gnp, Miller:2023kkd}~(solid) and projected~\cite{STE-QUEST:2022eww,Fedderke:2022ptm}~(dashed) bounds from DM searches~(blue); the latter strictly speaking does not apply to our case unless such scalar field is assumed to be dark matter.

\subsection{Pseudo-Scalar Coupling to Matter}

Pseudo-scalar couplings, as given by the second term in Eq.~\eqref{eq:interactions}, typically lead to spin-dependent effects due to the involved factor of $\gamma_{5}$. For non-relativistic fermions, the interaction can be written as~\cite{Graham:2013gfa},
\begin{equation}
    H\sim (\vec{\nabla}\phi)\cdot \vec{S}_{\psi},
\end{equation}
resulting in a torque on the spins,
\begin{equation}
    \vec{T}\sim (\vec{\nabla}\phi)\times\vec{S}_{\psi}.
\end{equation}

Due to the non-vanishing magnetic dipoles of the fermions a change in the fermion spin results in a change in the magnetic field which can be measured.

\subsubsection{Optical Magnetometry: GNOME}
Modern Optical Magnetometry can detect changes in ASD of atomic magnetic fields of the order of $\text{ASD}[B]\sim 10^{-15} \text{T}/\sqrt{\text{Hz}}$, with projected sensitivities down to $\text{ASD}[B]\sim 10^{-17} \text{T}/\sqrt{\text{Hz}}$~\cite{Budker_2007} or, equivalently, Zeeman-like energy splittings of order $\text{ASD}[\Delta E]\sim 10^{-21}\text{eV}/\sqrt{\text{Hz}}$~\cite{GNOME:2023rpz}. We consider as experimental setup the Global Network of Optical Magnetometers for Exotic physics searches (GNOME)~\cite{Pustelny:2013rza}.

More precisely, the effective magnetic field induced by a pseudo-scalar coupling is given by~\cite{Pospelov:2012mt}
\begin{equation}
    \mu \vec{B}_\text{eff.}\equiv \dfrac{\vec{\nabla}\phi}{f_\text{eff.}}\,,
\end{equation}
where $\mu$ is the nuclear magnetic moment of the material~\cite{Pirahmadian:2017ntv} and $f_\text{eff.}$ is the material-dependent effective interaction scale. In the case of $^3\text{He}$, the scale depends solely on the neutron coupling
\begin{align}
    &f^{-1}_\text{eff.}(^3\text{He})=\dfrac{1}{f_n}=\frac{g_P}{m_n}\,.
\end{align} 
We will consider it as a benchmark.

Neglecting $\mathcal{O}(1)$ corrections from the direction of arrival of the signal, the effective Zeeman-like energy splitting is therefore given by
\begin{equation}
    |\Delta E|\propto \dfrac{g_P}{m_n}|\partial_r \phi|\,.
\end{equation}

\begin{figure}[]
    \centering
    \subfigure[{}\label{fig:optical-ASD-KG}]{\includegraphics[width=0.49\textwidth]{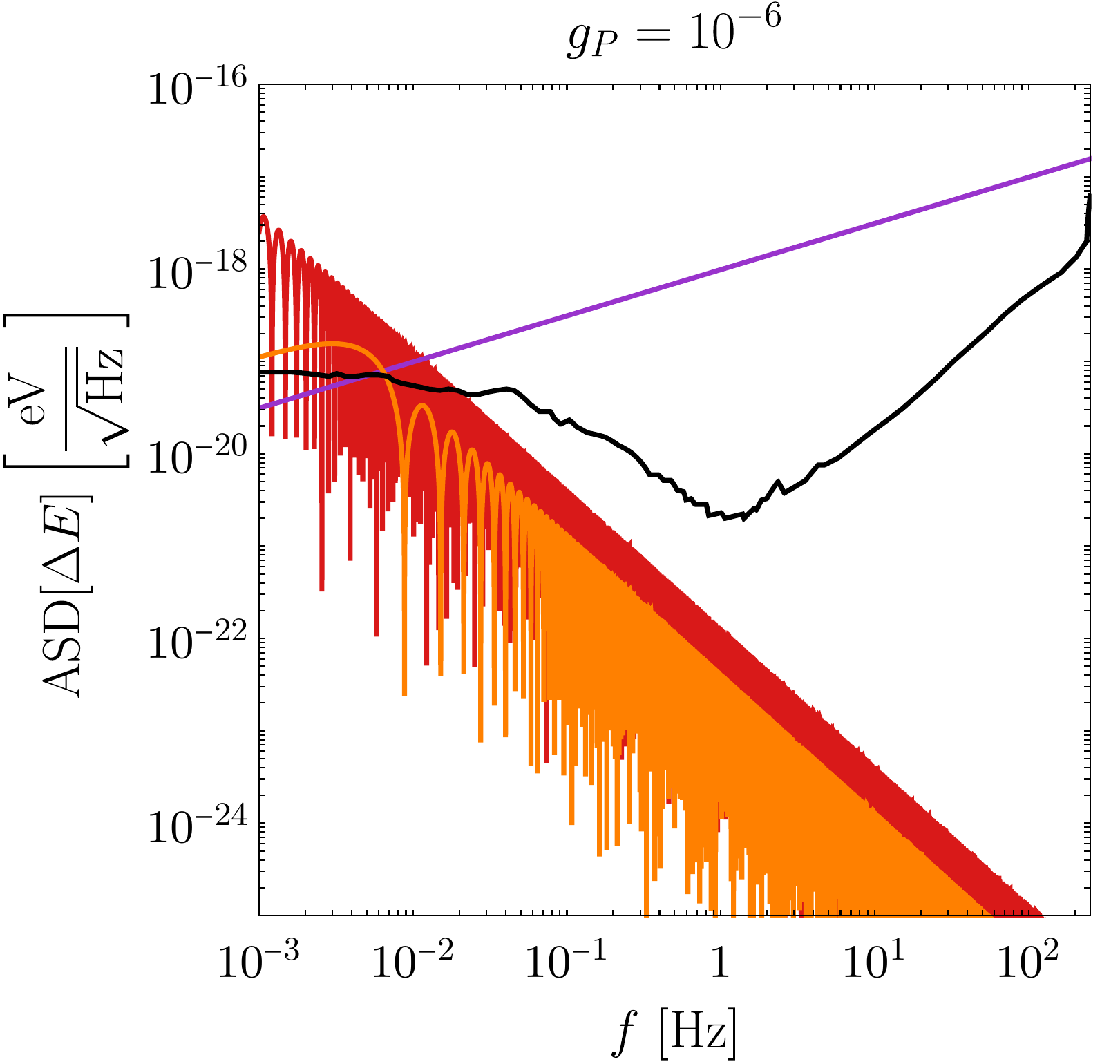}}
    \subfigure[{}\label{fig:optical-ASD-CS}]{\includegraphics[width=0.49\textwidth]{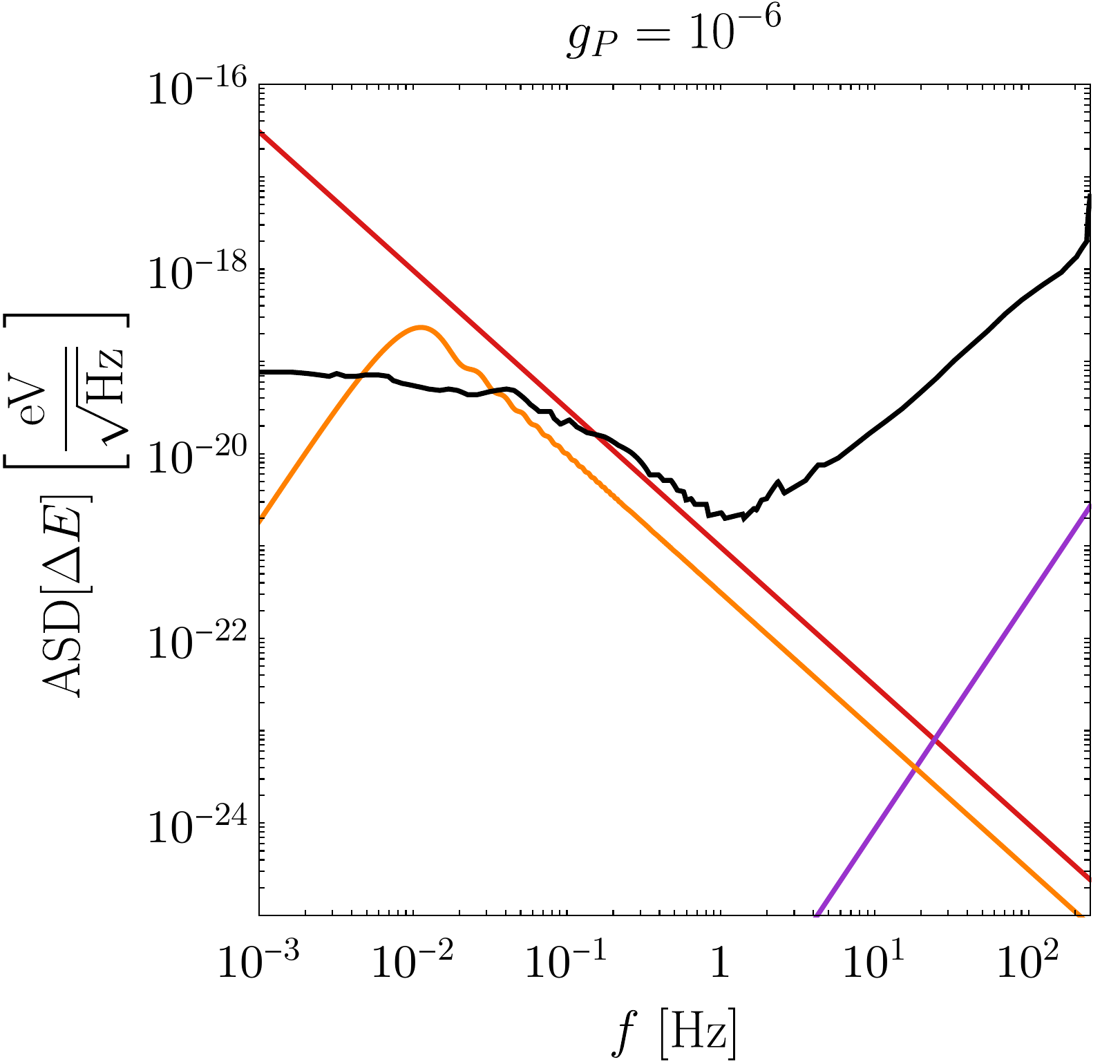}}
    \caption{ASD of the Zeeman-like effect induced by the signals (red giant~(\redG{red}), supernova~(\orangeG{orange}) and neutron star~(\purpleG{purple})) compared to the sensibility of the GNOME Krakow Station (black)~\cite{GNOME:2023rpz} for a YL-~(left) and CS-field~(right). The coupling was fixed to $g_P=10^{-6}$ and $E=10^{-1}\, M_\odot$.}
    \label{fig:Optical-Kracow}
\end{figure}
A comparison of the induced signal with the noise spectrum of the GNOME Krakow Station~\cite{GNOME:2023rpz} can be seen in Fig.~\ref{fig:Optical-Kracow}.
    
Integrating over the full available range of frequencies and imposing $\text{SNR} \leq 1$ we obtain

\begin{align}
    &\label{eq:B-OPT-1} g^\text{Red Giant}_\text{P}\lesssim \begin{Bmatrix}
       2\times 10^{-11}\,, & \text{YL}\\
       2\times 10^{-12}\,, & \text{CS}
    \end{Bmatrix}\times\left(\sqrt{\dfrac{10^{-1}\,M_\odot}{E}}\dfrac{r}{500\,[\text{ly}]}\right)\,,\\
    &\label{eq:B-OPT-2} g^\text{SN1987A}_\text{P}\lesssim \begin{Bmatrix}
       2\times 10^{-7}\,, & \text{YL}\\
       1\times 10^{-7}\,, & \text{CS}
    \end{Bmatrix}\times\left(\sqrt{\dfrac{10^{-1}\,M_\odot}{E}}\dfrac{r}{1.7\times 10^{5}\,[\text{ly}]}\right)\,,\\
    &\label{eq:B-OPT-3} g^\text{NS}_\text{P}\lesssim \begin{Bmatrix}
       8\times 10^{-10}\,, & \text{YL}\\
       5\times 10^{-2}\,, & \text{CS}
    \end{Bmatrix}\times\left(\sqrt{\dfrac{10^{-1}\,M_\odot}{E}}\dfrac{r}{400\,[\text{ly}]}\right)\,,\\
    &\,\,\,\quad\nonumber\lesssim\begin{Bmatrix}
       3\times 10^{-4}\,, & \text{YL}\\
       2\times 10^{4}\,, & \text{CS}
    \end{Bmatrix}\times\left(\sqrt{\dfrac{10^{-1}\,M_\odot}{E}}\dfrac{r}{1.3\times 10^{8}\,[\text{ly}]}\right)\,.
\end{align}

\begin{figure}[H]
    \centering
    \includegraphics[width=0.9\textwidth]{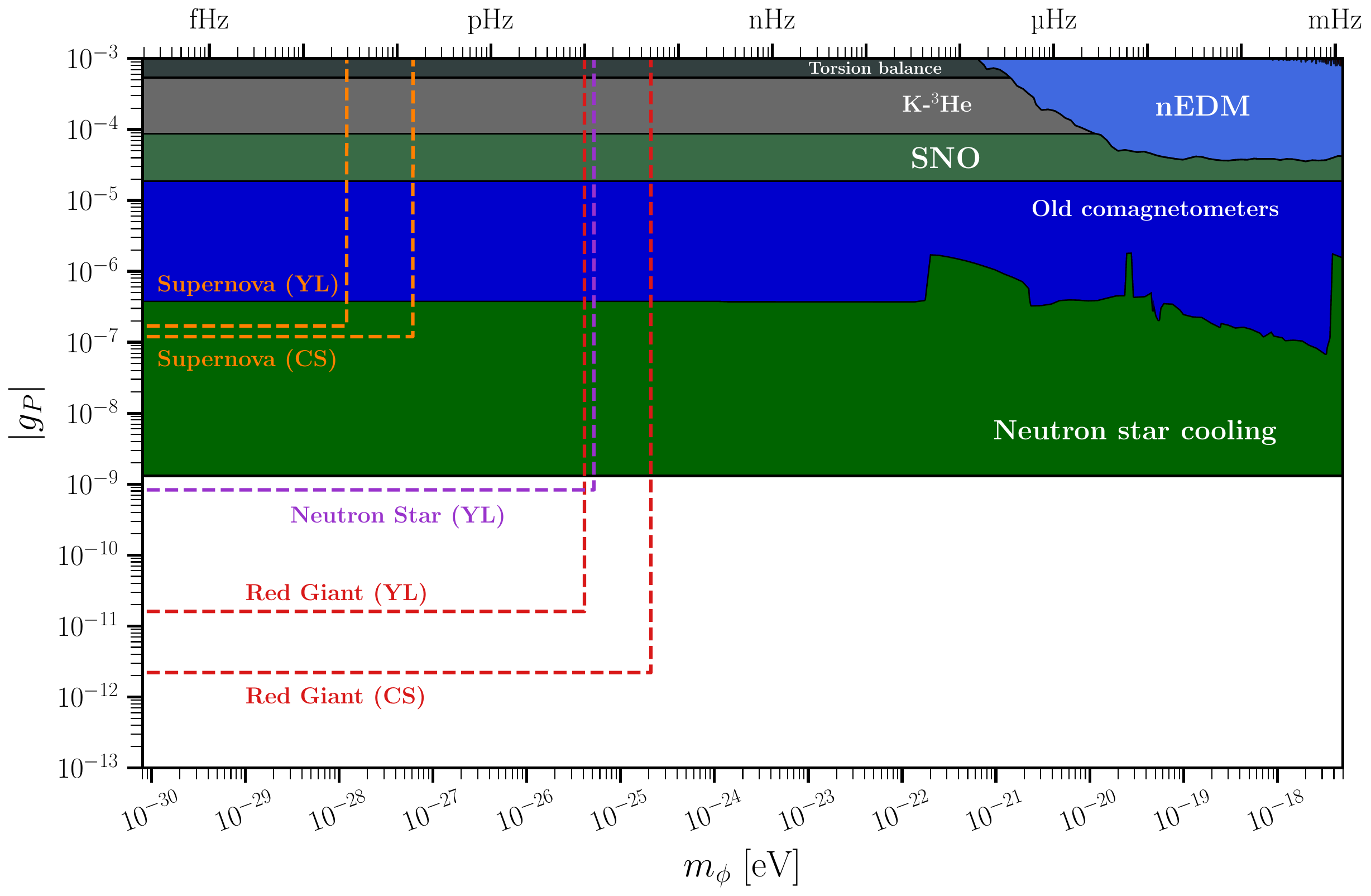}
    \caption{Sensitivities from Eqs.~\eqref{eq:B-OPT-1}, \eqref{eq:B-OPT-2} and \eqref{eq:B-OPT-3} adopting $E=10^{-1}\,M_\odot$ on the pseudo-scalar neutron coupling $|g_P|$ for a YL- and a CS-field. We also report bounds from astrophysical sources~(green), laboratory tests of the equivalence principle~(grey) and current bounds from DM searches~(blue). Plot adapted from Refs.~\cite{OHare:2020wah,ohare}. The bounds are taken from Refs.~\cite{Adelberger:2006dh, Vasilakis:2008yn,Bhusal:2020bvx, Buschmann:2021juv,Bloch:2019lcy, Centers:2019dyn}.}
    \label{fig:coupling-bounds-P}
\end{figure}

The resulting sensitivities from Eqs.~\eqref{eq:B-OPT-1}, \eqref{eq:B-OPT-2} and \eqref{eq:B-OPT-3} adopting $E=10^{-1}\,M_\odot$ can be seen in Fig.~\ref{fig:coupling-bounds-P} for the red giant~(\redG{red}), supernova~(\orangeG{orange}) and neutron star~(\purpleG{purple}) case. The neutron star case for the CS-signal is not shown as the sensitivity is not significant. We compare the results with current bounds from the literature which include laboratory tests~\cite{Adelberger:2006dh, Vasilakis:2008yn} in grey, astrophysical bounds~\cite{Bhusal:2020bvx, Buschmann:2021juv} in green and DM searches~\cite{Bloch:2019lcy, Centers:2019dyn} in blue; the latter bounds do not necessarily apply to our case.

\subsection{Two-Photon Coupling}
We can also consider couplings directly to the electromagnetic field, e.g. a typical coupling for an axion-like particle, cf. e.g.~\cite{Sikivie:1983ip,Georgi:1986df},
\begin{equation}
    \mathcal{L}\supset-\dfrac{ g_{\gamma}}{4}\phi F^{\mu\nu}\Tilde{F}_{\mu\nu}= g_{\gamma}\phi \Vec{E}\cdot\Vec{B}\,.
\end{equation}
The presence of a non-vanishing $\phi$ modifies Ampere's Law generating an effective current~\cite{Chaudhuri:2018rqn}
\begin{equation}
    \Vec{J}\approx -g_{\gamma}\left(\Vec{B}_b\dfrac{\partial\phi}{\partial t}-\Vec{E}_b\times\Vec{\nabla}\phi\right)\,,
\end{equation}
where $\vec{E}_b,~\vec{B}_b$ are the electric and magnetic background fields, respectively. 
This can then be detected in suitable experiments.

In the following, we will consider two versions of DMRadio~\cite{DMRadio:2022pkf} originally aimed at searching for axion and ALP dark matter~\cite{Preskill:1982cy, Abbott:1982af, Dine:1982ah,Arias:2012az} that are sensitive to the relatively low frequencies we are interested in.

The induced current in DMRadio obeys a harmonic-oscillator-type differential equation
\begin{align}
    & L\ddot{I}+R\dot{I}+\dfrac{I}{C} = \dot{V}_\phi(t)\,,  &\ddot{I}+\dfrac{\omega_0}{Q}\dot{I}+\omega_0^2 I = \dfrac{\dot{V}_\phi(t)}{L}\,,
\end{align} 
where $V_\phi(t)$ is the external voltage, the resonant angular frequency is given by $\omega_0^2=(2\pi f_0)^2=1/LC$ and the quality factor reads $Q=\sqrt{L/CR^2}$.
The measurable quantity of interest is the current $I$.

\subsubsection{DMRadio-m$^3$}
Let us consider DMRadio-m$^3$~\cite{DMRadio:2022pkf, DMRadio:2023igr}. The experiment uses a static solenoid magnet to
drive the induced effective current $\vec{J}$ along the axis of a coaxial inductive pickup~(PU), which then couples to a tunable
capacitor to form an LC resonator circuit. The reported quality factor, background magnetic field and inductance PU of $Q>10^5$, $B_b>4$~T and $L_\text{PU}\lesssim 200$~nH are reported as an achievable goal, and we will therefore use such values as a benchmark.  The resonance frequency is tuned in the range $[5-200]~\text{MHz}$ by varying the capacitor in the range $10~\text{pF}\lesssim C \lesssim 5~\text{nF}$. The estimated damping times are of order
\begin{equation}
    t_\text{damping}\sim\frac{Q}{2\pi f_0}\sim (10^{-4}-10^{-3})~\text{s}\,.
\end{equation}
This is a much shorter time and smaller than the expected measurement time, in contrast to the case of the torsion pendulum.

The transient signal induces a potential $V_\phi$ that can be computed using Faraday's Law to be
\begin{equation}
    V_\phi=-\frac{\diff{\Phi_B}}{\diff{t}}\,.
\end{equation}
The magnetic field induced into the main circuit can be computed via Biot-Savart law
\begin{equation}
    \vec{B}(\Vec{r})=\frac{\mu_0}{4\pi}\int_V \diff\Vec{r}' \frac{\Vec{J}\times (\Vec{r}-\Vec{r}')}{|\Vec{r}-\Vec{r}'|^3}\,.
\end{equation}
As the induced current is proportional to the background magnetic field, $\Vec{B}_b$, we obtain
\begin{equation}
    |\Phi_B|=\sqrt{2\mu_0}c_\text{PU}|\Vec{J}|\left(V_\text{PU}^{5/3}L_\text{eff.}\right)^{1/2}\,,
\end{equation}
and thus
\begin{equation}
    |V_\phi(t)|=-\sqrt{2\mu_0}c_\text{PU}|\partial_t\Vec{J}|\left(V_\text{PU}^{5/3}L_\text{eff.}\right)^{1/2}\,,
\end{equation}
where $V_\text{PU}$ is the pickup volume, $c_\text{PU}$ is a constant that depends on the geometry of the apparatus, ideally $L_\text{eff}\approx L_\text{PU}$ and $V_\text{PU}\approx 2~\text{m}^3$. We will employ $c_\text{PU}=0.15$ as DM-radio-m$^3$ reports values in the range $c_\text{PU}\approx 0.1-0.2$. Finally, the ASD induced current reads
\begin{equation}
    |\widetilde{I}| = \frac{|\widetilde{\dot{V}}_\phi|}{\omega\,|Z(\omega)|} \,,
\end{equation}
being $Z(\omega)$ the circuit impedance and $\widetilde{V}$ the Fourier-transform of the induced potential and the squared impedance reads
\begin{equation}
    |Z(\omega)|^2=\dfrac{L^2\omega_0^2}{Q^2}\left[1+Q^2\dfrac{\omega_0^2}{\omega^2}\left(1-\dfrac{\omega^2}{\omega_0^2}\right)^2\right]\,.
\end{equation}
All in all, the signal enters the expression with the current with its third-time derivative.
The current thermal noise power spectrum is given by
\begin{align}
    &\label{eq:thermal-noise}S_T(\omega)=4\hbar\omega\left(n_T+\dfrac{1}{2}\right)\dfrac{\Re{Z(\omega)}}{|Z(\omega)|^2}\,,  &n_T=\dfrac{1}{\exp{(\hbar\omega)/(k_BT)}-1}\,.
\end{align}
The target of DMRadio-m$^3$ is a temperature of $T=20~\text{mK}$, which will be used as a benchmark. We have not included the amplifier imprecision noise and the amplifier backaction noise.

\begin{figure}[]
\centering
    \subfigure[{}\label{fig:resonant-cavity-ESD-KG}]{\includegraphics[width=0.48\textwidth]{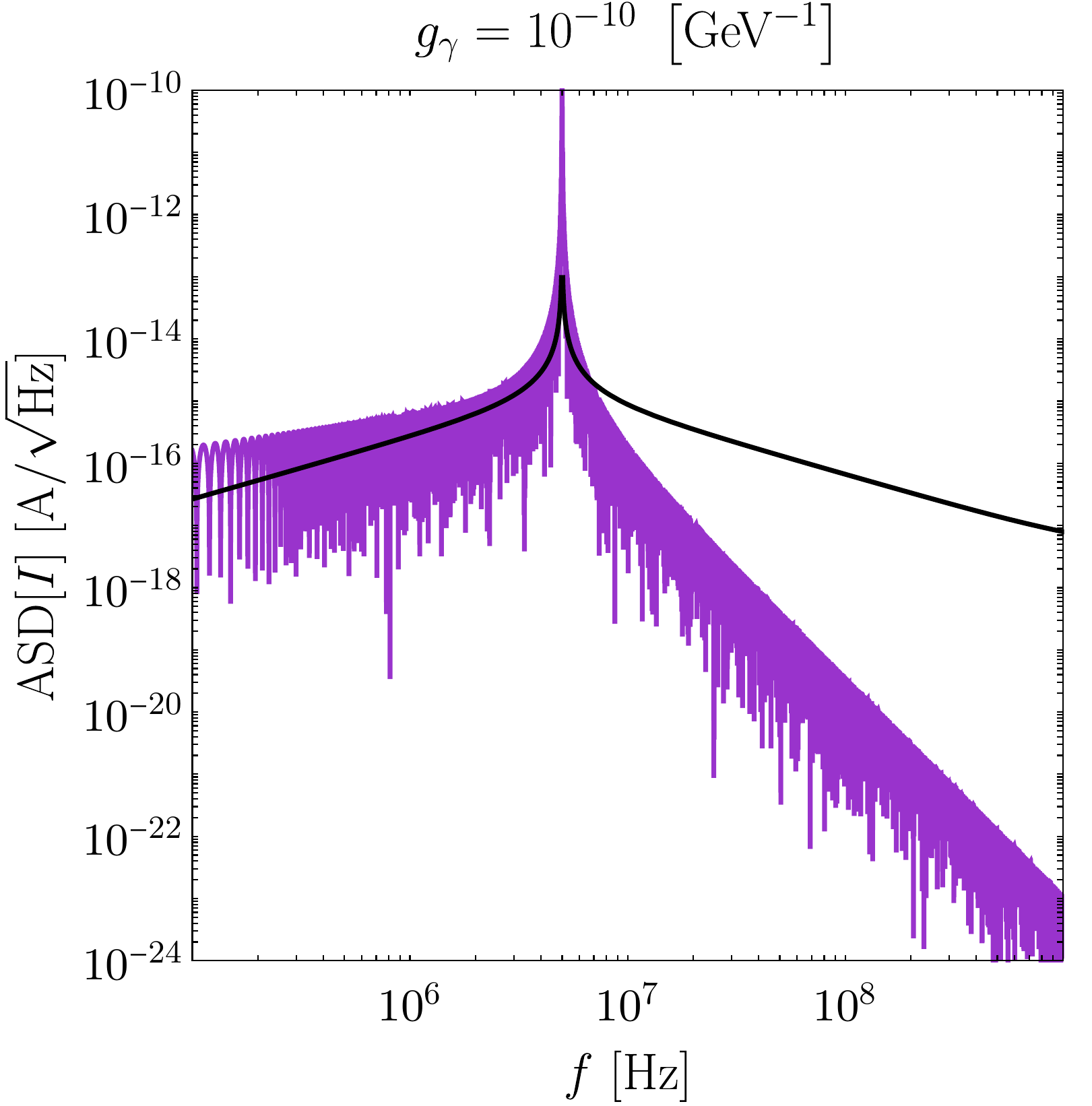}}
    \subfigure[{}\label{fig:resonant-cavity-ESD-C4}]{\includegraphics[width=0.48\textwidth]{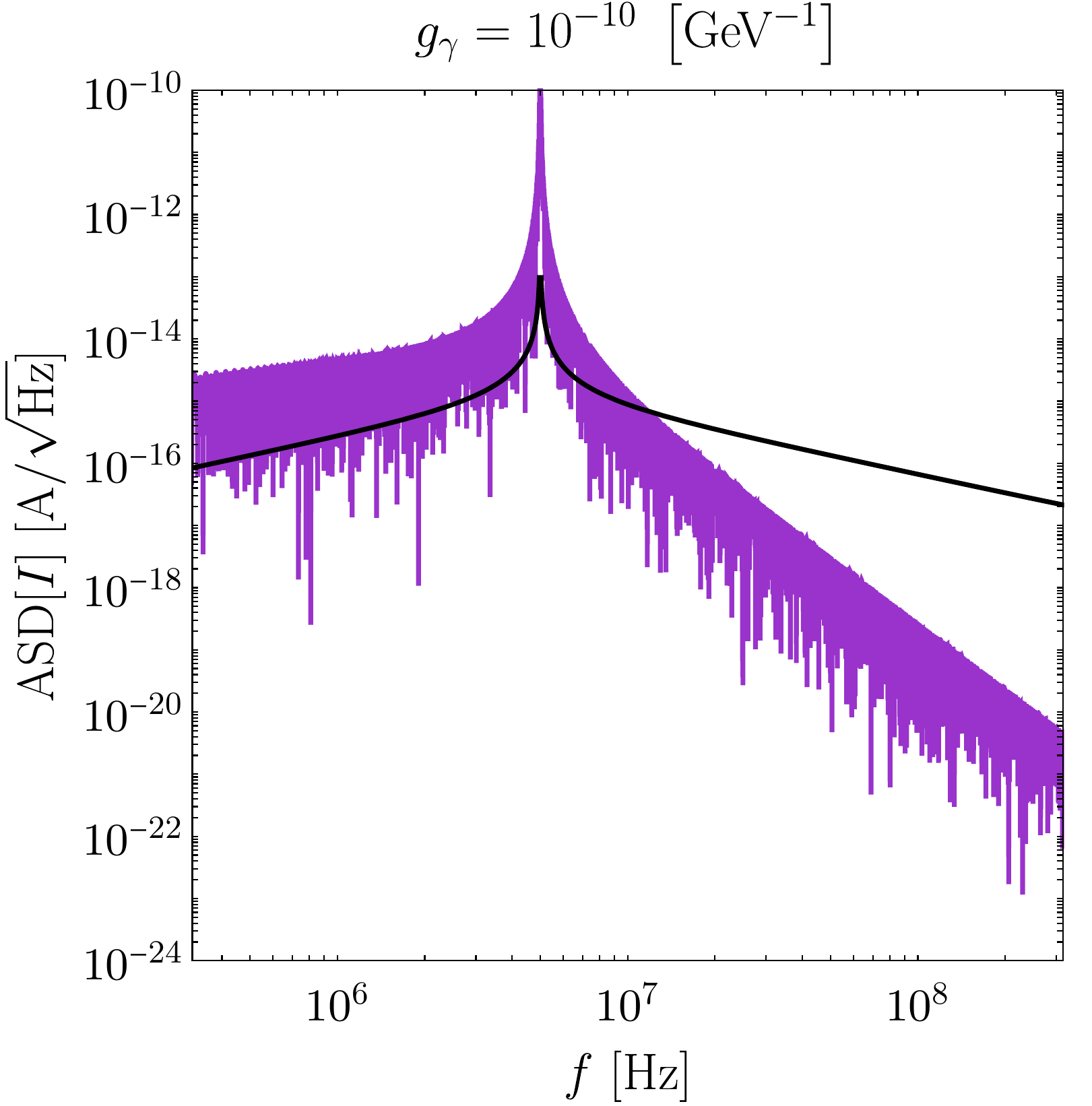}}
    \caption{Energy spectral density of the YL-~(left) and CS~(right) signals (\redG{red}), supernova~(\orangeG{orange}) and neutron star~(\purpleG{purple})). The coupling was fixed to $g_{\gamma}=10^{-10}~\GeV^{-1}$ and $E=10^{-1}\, M_\odot$. The resonant frequency was fixed to $f_0=5~\text{MHz}$. The thermal noise of Eq.~\eqref{eq:thermal-noise} is shown in black. Only the NS signal appears as the Supernova is strongly suppressed by its large radius.}
    \label{fig:resonant-cavity-ESD}
\end{figure}

Finally, noticed that the induced potential is proportional to the third time derivative of $\phi(r,t)$
\begin{align}
    &\left|\widetilde{\dddot{\phi}}_\text{YL}\right|=\sqrt{\frac{15\, ER}{\pi }}\frac{| \sin (\omega R)|}{2 r R^3 \omega}\propto\begin{cases}
         1 & \omega R \ll 1\,,\\
         (\omega R)^{-1} & \omega R \gg 1\,.
    \end{cases}\\
    &\left|\widetilde{\dddot{\phi}}_\text{CS}\right|=\sqrt{\frac{210\, ER}{\pi }}\frac{|\omega R   \sin (\omega R  )+4 \cos (\omega R  )-4|}{r R^4 \omega ^2}\propto\begin{cases}
        1 & \omega R \ll 1\,,\\
         (\omega R)^{-1} & \omega R \gg 1\,.
    \end{cases}
\end{align}
Such asymptotics can be seen explicitly in Fig.~\ref{fig:resonant-cavity-ESD}.
The signal of the supernova is strongly suppressed to the NS one due to its large radius. Considering only thermal noise, one can obtain the following bounds from the SNR
\begin{align}
    &\label{eq:B-RC-m3-1}g^\text{NS}_{\gamma}\lesssim \begin{Bmatrix}
       6\times 10^{-10}\,, & \text{YL}\\
       8\times 10^{-11}\,, & \text{CS}
    \end{Bmatrix}\times\left(\sqrt{\dfrac{10^{-1}\,M_\odot}{E}}\dfrac{r}{400\,[\text{ly}]}\right)\,[\GeV^{-1}]\,,\\
    &\,\,\,\quad\nonumber\lesssim\begin{Bmatrix}
       2\times 10^{-4}\,, & \text{YL}\\
       3\times 10^{-5}\,, & \text{CS}
    \end{Bmatrix}\times\left(\sqrt{\dfrac{10^{-1}\,M_\odot}{E}}\dfrac{r}{1.3\times 10^{8}\,[\text{ly}]}\right)\,[\GeV^{-1}]\,.
\end{align}

\subsubsection{DMRadio-50L}
The sensitivity to the type of transients we are interested in can be significantly improved by using an experiment with goo sensitivity in the kHz range. A candidate experiment for such a frequency range is DMRadio-50L. The experiment aims to scan the range of frequencies in the band $5~\text{kHz} \,-\, 5~ \text{MHz}$ Details about its components can be found in Ref.~\cite{DM50}.  The test benchmark point of the resonator uses $C=220$~pF and $L=500~\mu$H, yielding a resonant frequency of $f\approx 480$~kHz. The quality factor is reported to be $Q\sim 10^6$, which will be used as a benchmark. Furthermore, we employ a peak background magnetic field of $B=1$~T and a volume of $V=50\text{L}=0.05~\text{m}^3$. As for the case study of DMRadio-m$^3$, we will set the thermal noise temperature to $T=20~\text{mK}$. 

Following the same procedure described before, the following bounds can be obtained
\begin{align}
    &\label{eq:B-RC-50-1}g^\text{NS}_{\gamma}\lesssim \begin{Bmatrix}
       4\times 10^{-8}\,, & \text{YL}\\
       4\times 10^{-9}\,, & \text{CS}
    \end{Bmatrix}\times\left(\sqrt{\dfrac{10^{-1}\,M_\odot}{E}}\dfrac{r}{400\,[\text{ly}]}\right)\,[\GeV^{-1}]\,,\\
    &\,\,\,\quad\nonumber\lesssim\begin{Bmatrix}
       1\times 10^{-2}\,, & \text{YL}\\
       1\times 10^{-3}\,, & \text{CS}
    \end{Bmatrix}\times\left(\sqrt{\dfrac{10^{-1}\,M_\odot}{E}}\dfrac{r}{1.3\times 10^{8}\,[\text{ly}]}\right)\,[\GeV^{-1}]\,.
\end{align}
Despite the better frequency match the bounds are slightly weaker than those derived from DMRadio-m$^3$. This is due to the much smaller pick-up volume and magnetic field. Moreover, the optimal frequency for a NS of $f_\text{NS}\sim 5$~kHz is still on the edge of the frequency band of DMRadio-50L. If the resonant frequency was tuned to such value an improvement of the bound by a factor of $10-10^2$ is expected. 

\begin{figure}[H]
    \centering
    \includegraphics[width=0.9\textwidth]{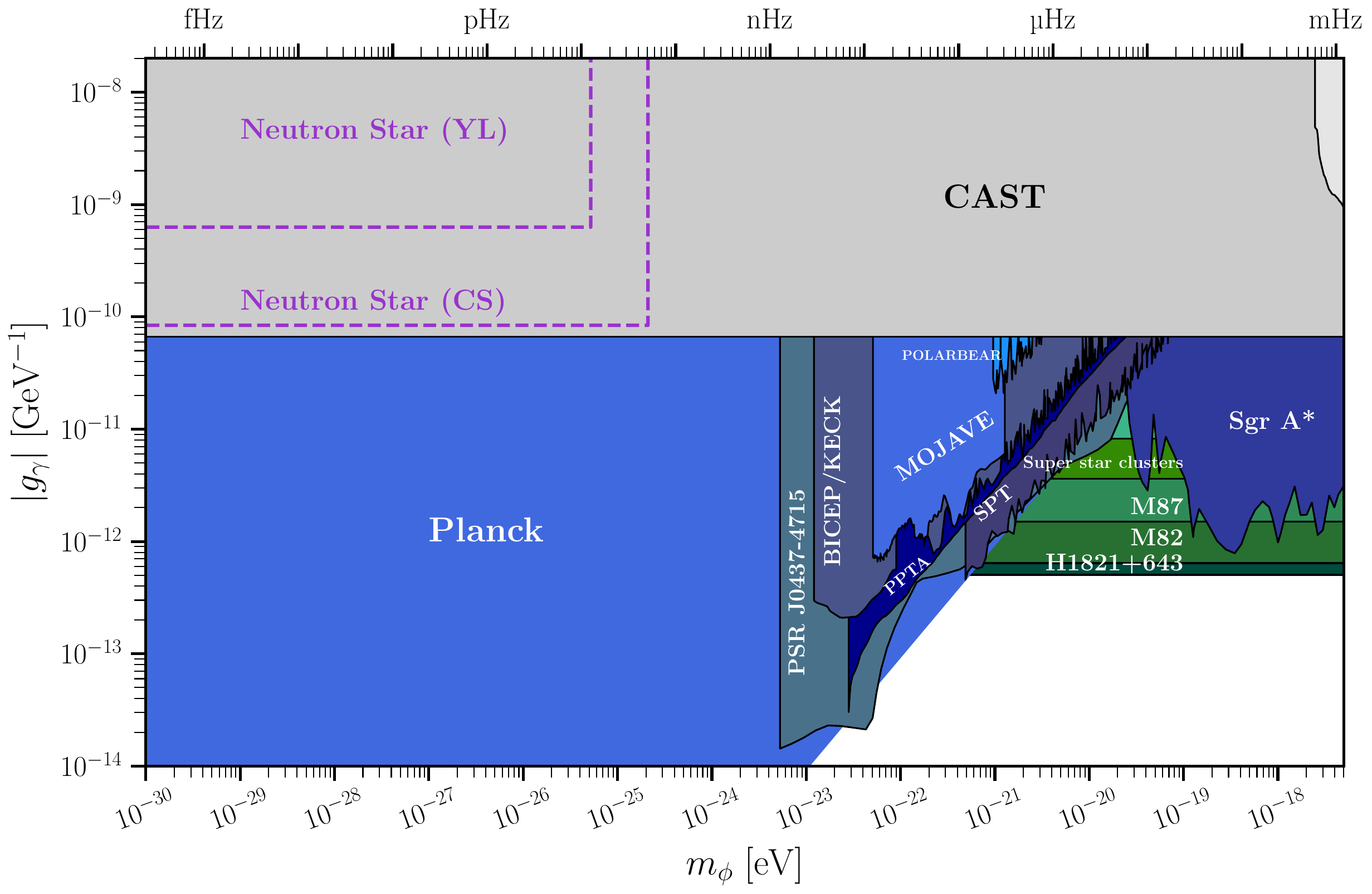}
    \caption{Sensitivities on the two-photon coupling $|g_{\gamma}|$ reported in Eq.~\eqref{eq:B-RC-m3-1} for a YL- and a CS-field adopting $E=10^{-1}\,M_\odot$. We also show bounds from astrophysical sources~(green), DM bounds~(blue) and laboratory tests~(grey). Plot adapted from Refs.~\cite{OHare:2020wah,ohare}.  The bounds shown in the plot are taken from Refs.~\cite{CAST:2017uph,Marsh:2017yvc, Dessert:2020lil, Ning:2024eky,Ivanov:2018byi,Fedderke:2019ajk,Caputo:2019tms,Yuan:2020xui,Fujita:2020aqt,BICEPKeck:2021sbt,Castillo:2022zfl,SPT-3G:2022ods,POLARBEAR:2023ric,POLARBEAR:2024vel}.}
    \label{fig:photonsumm}
\end{figure}

The best achievable sensitivity derived in Eq.~\eqref{eq:B-RC-m3-1} is shown in Fig.~\ref{fig:photonsumm} for the Neutron Star case; the results for the other sources are far from being competitive and do not appear in the shown parameter space. We compare such results with current bounds from the literature which include laboratory testes~\cite{CAST:2017uph} in grey and astrophysical bounds~\cite{Marsh:2017yvc, Dessert:2020lil, Ning:2024eky} in green. Even though they may not apply in our case, we report DM bounds~\cite{Ivanov:2018byi,Fedderke:2019ajk,Caputo:2019tms,Yuan:2020xui,Fujita:2020aqt,BICEPKeck:2021sbt,Castillo:2022zfl,SPT-3G:2022ods,POLARBEAR:2023ric,POLARBEAR:2024vel} in blue for completeness.

\section{Conclusions}\label{sec:conclusions}
In the search for very light and feebly interacting particles, usually, three types of interesting signals are considered. Static fields that are sourced from bodies of ordinary matter coupled to the new light fields/particles and that can be detected, e.g. via forces between test objects (cf., e.g.~\cite{Adelberger:2003zx} for a review). Oscillating fields and corresponding signals, with long persistence and also long coherence times, often arise if the new particles are dark matter (cf., e.g.~\cite{Antypas:2022asj} for a recent overview). Finally, there are transient signals, where the interaction with Earth-bound experiments is concentrated in a relatively short time and typically does not have a monochromatic structure.
While interesting works in this direction exist~\cite{Pospelov:2012mt,Jaeckel:2016jlh,Roberts:2017hla,Roberts:2018xqn,Dailey:2020sxa,Afach:2021pfd,Cuadrat-Grzybowski:2024uph,Eby:2021ece,GNOME:2023rpz,Arakawa:2023gyq,Maseizik:2024qly} the latter have received far less attention. In this context we want to discuss an interesting class of potential sources for this type of signals: originally static field configurations sourced by a large object of ordinary matter that is ``destroyed'' in a violent event where the source ``vanishes'' notably by forming a black hole. Examples are black hole-forming supernovae or neutron star mergers.
Providing potential new sources of transient signals we want to stimulate experiments to also consider this type of signals in their analyses.

As in this paper, our goal was mainly phenomenological, we focused on a simple modelling of potential signals and their main characteristic scales and left a first principle calculation to future work. We therefore acknowledge the existence of rather large modelling uncertainties (discussed in more detail in section~\ref{subsection:modeling}).  Notably, this includes gravitational effects, and importantly the scalar field evolution during the ``collapse''. If the evolution is fast, as one may expect, e.g. in the case of weak self-interactions this can potentially modify our sensitivity by orders of magnitude.
Our estimates for the sensitivity, frequencies and signal shapes should therefore be taken as only indicative of what could be achieved and where to look.
They are summarized in Figs.~\ref{fig:coupling-bounds} (scalar coupling), \ref{fig:coupling-bounds-P} (pseudo-scalar coupling) and \ref{fig:photonsumm} (two-photon coupling) for the different types of coupling we considered. We stress that these are only examples and a wider range of couplings (e.g. to vectors) could be considered even with already existing experiments. Notably, relatively nearby sources such as a Betelgeuse going supernova may already provide discovery potential, encouraging a dedicated analysis for such signals. Experiments that are more adapted to transient signals and targeting signal frequency ranges more suitable for this type of signal could certainly increase the sensitivity significantly. Although we have not made use of this in our analysis we note that the cosmic events we consider may be visible also in other observables, e.g. gravitational waves for mergers or optical or neutrino signals for supernovae. Looking for coincidences could be used to improve the sensitivity.

Finally, we also note that the events we considered, as well as other types of transient events, e.g.~\cite{Dailey:2020sxa,Eby:2021ece,Arakawa:2023gyq,GNOME:2023rpz,Maseizik:2024qly} may also lead to a stochastic background when many events are accumulated over the cosmos\footnote{In spirit this is similar to, for example, the stochastic backgrounds of neutrinos, axion and axion-like particles, created by supernovae events in the cosmic history, cf.~e.g.~\cite{Beacom:2010kk,Raffelt:2011ft,Calore:2020tjw}. However, the features, notably energies would be very different. This could even be considered a fourth type of signal for very low mass feebly interacting particles.}. This could be an interesting target for future studies.

\section*{Acknowledgements}
AdG thanks the
Institute for Theoretical Physics of the University Heidelberg for the warm hospitality during part of the realization of this work. Our work is supported by the European Union’s Horizon 2020 Marie Sklodowska-Curie grant agreement No 860881-HIDDeN. AdG acknowledges the support of the Spanish Agencia Estatal de Investigacion through the grant “IFT Centro de Excelencia Severo Ochoa CEX2020-001007-S”. This
article/publication is based upon work from COST Action COSMIC WISPers CA21106,
supported by COST (European Cooperation in Science and Technology).

\appendix

\section{Solutions to the Klein-Gordon Equation}
\label{app:Klein-Gordon}

\subsection{KG Green's Function and Applications}
The Green's function of the KG equation is well known for the $m_\phi=0$ case, but it is harder to apply in the massive case. Let us review what it looks like. For a review of the method see e.g. Ref.~\cite{Peskin:1995ev}.
We are interested in solving
\begin{equation}
    (\Box+m_\phi^2)G(x)=\delta(x)\,,
\end{equation}
where the delta acts on all the four components of $x$. Using the Fourier transform properties
\begin{align}
    &G(x)=\int\dfrac{\diff^4k}{(2\pi)^2}e^{ikx}\widetilde{G}(k)\,,\\
    &\delta(x)=\int\dfrac{\diff^4k}{(2\pi)^4}e^{ikx}\,,
\end{align}
one can show that
\begin{equation}
    G(x)=-\int\dfrac{\diff^4k}{(2\pi)^4}\dfrac{e^{ikx}}{k^2-m_\phi^2}\,.
\end{equation}
By performing the time integration in the lower $k^0$ contour, we obtain the retarded-Green's function
\begin{equation}
\label{eq:Green-Function}
    G_r(x)=2\Theta(t)\int\limits\dfrac{\diff^3k}{(2\pi)^3 2\omega_k}\,e^{-i\Vec{k}\cdot\Vec{x}}\sin(\omega_k t)=\dfrac{\Theta(t)}{|\Vec{x}|}\dfrac{1}{2\pi^2}\int\limits_0^\infty\diff{k}\,\dfrac{k}{\omega_k}\sin(k |\vec{x}|)\sin(\omega_k t)\,,
\end{equation}
where $\omega_k^2\equiv m_\phi^2+k^2$\,.

\subsubsection{Application to Point-Like Source}
We now apply the result of Eq.~\eqref{eq:Green-Function} to the case of a point-like source that is suddenly turned off at $t=0$. 
Given a term $\phi J(x) $ in the Lagrangian, the solution of the inhomogeneous KG equation is given by
\begin{equation}
    \phi(x)=\int\diff^4y \,G(x,y)J(y)\,.
\end{equation}
In our case, the current can be written as
\begin{equation}
    J(x)=g\Theta(-t)\delta^{(3)}(\Vec{x})\,.
\end{equation}
After regularizing the integral in $y^0$ via a $e^{\epsilon y^0}$ factor, the integration yields
\begin{equation}
    \phi(x)=\dfrac{g}{2\pi^2|\Vec{x}|}\int\limits_0^\infty \diff{k}\,\dfrac{k}{\omega_k^2}\sin(k|\Vec{x}|)\cos((t-\widetilde{x}^0)\omega_k)\,,
\end{equation}
where $\widetilde{x}^0\equiv \min\{0,t\}=t\Theta(-t)$.
For the case $t\leq 0$, the solution reads
\begin{equation}
    \phi(t\leq 0,\Vec{x})=\dfrac{g}{4\pi}\dfrac{e^{-m_\phi|\vec{x}|}}{|\vec{x}|}\,,
\end{equation}
which is the typical static solution for a static source.

\subsubsection{Application to a Spherical Source of Size $R$}
We consider here a spherical source of radius $R$. The source term can be written as
\begin{equation}
    J(x)=\dfrac{3g}{4\pi R^3}\Theta(-t)\Theta(R-|\vec{x}|)\,.
\end{equation}
Similarly to the previous section, the time integration can be carried out quite easily. The spatial integration on the contrary is much more involved. Respectively
\begin{align}
    &\int_\mathbb{R}\diff{y}^0 \sin(\omega_k(t-y^0))\Theta(t-y^0)\Theta(-y^0)=\dfrac{\cos(\omega_k(t-\widetilde{x}^0))}{\omega_k}\,,\\
    & \int_{\mathbb{R}^3}\diff^3y\, \dfrac{\sin(k|\vec{x}-\vec{y}|)}{|\vec{x}-\vec{y}|}\Theta(R-|\vec{y}|)=\dfrac{4\pi R}{|\vec{x}|}\dfrac{\sin(k|\vec{x}|)}{k^2}\left[\dfrac{\sin(kR)}{k R}-\cos(kR)\right]\,.
\end{align}
The solution is then given by
\begin{equation}
\label{eq:field_R-app}
    \phi(x)=\dfrac{3g}{2\pi^2R^2|\vec{x}|}\int\limits_0^\infty \diff{k}\,\dfrac{1}{\omega_k^2}\dfrac{\sin(k|\Vec{x}|)}{k}\cos((t-\widetilde{x}^0)\omega_k)\left[\dfrac{\sin(k R)}{kR}-\cos{(kR)}\right]\,.
\end{equation}

In the massless limit, the result can be computed exactly. For $t<0$ and $r>R$, the solution is static and gives the usual
\begin{equation}
    \phi(t\leq 0)=\dfrac{g}{4\pi r}\,.
\end{equation}
For $t>0$, it gets more involved
\begin{equation}
\label{eq:field-r>R}
    \phi(r\geq R)=\dfrac{g}{4\pi r}\times \begin{cases}
    1\,,& t\in [0,r-R]\\\
    -\dfrac{(r-2R-t)(r+R-t)^2}{4R^2}\,,&t\in[r-R,r+R]\\
    0\,,&t\in[r+R,+\infty)
    \end{cases}\,.
\end{equation}
Similarly, for $t>0$ we can extend the solution of Eq.~\eqref{eq:field-r>R} to the case $r\leq R$ and find
\begin{equation}
\label{eq:field-r<R}
    \phi(r\leq R)=-\dfrac{g}{8\pi R^3}\times \begin{cases}
    \left(r^2-3R^2+3t^2\right)\,,& t\in [0,R-r]\\\
    \dfrac{1}{2r}(r-2R-t)(r+R-t)^2\,,&t\in[R-r,R+r]\\
    0\,,&t\in[R+r,+\infty)
    \end{cases}\,.
\end{equation}

\subsection{Force and Energy for a Spherical Source}
\subsubsection{Force}
We focus on the results for $r\geq R$. The force that is exerted is proportional to the gradient of the field. We conventionally take one of the basis axes aligned with the line that joins the source centre and the particle on whom the force acts. The force is then given by
\begin{equation}
    F(r,t)=-\partial_r\phi\,.
\end{equation}
The expression for the force thus reads
\begin{equation}
\label{eq:force-generic}
    F(r,t)=\dfrac{3g}{2\pi^2 r R^2}\int\limits_0^\infty\diff{k}\,\dfrac{\cos((t-\widetilde{x}^0)\omega_k)}{\omega_k^2}\mathcal{F}(kr)\mathcal{F}(k
R)\,,
\end{equation}
where we have defined
\begin{equation}
\label{eq:def-F}
    \mathcal{F}(x)\equiv \dfrac{\sin(x)}{x}-\cos(x)\,.
\end{equation}
The solution cannot be computed exactly but for the massless case; it reads for $r\geq R$
\begin{equation}
\label{eq:force}
    F(r,t)=\dfrac{g}{4\pi r}\times \begin{cases}
    \dfrac{1}{r}\,,& t\in [0,r-R]\\\
  \dfrac{t^3-3t(r^2+R^2)+2(r^3+R^3)}{4r R^3}\,,&t\in[r-R,r+R]\\
    0\,,&t\in[r+R,+\infty)
    \end{cases}\,.
\end{equation}
The result is particularly interesting as it explicitly shows an effective $1/r$ dependence of the amplitude during the time-dependent transient.

\subsubsection{Energy}
Let us also briefly consider the energy contained in the configuration.
The Hamiltonian density for the field is given by
\begin{equation}
\label{eq:energy}
    \mathcal{H}=\dfrac{1}{2}\left[(\dot{\phi})^2+(\phi')^2+m_\phi^2\phi^2-2\phi J(x)\right]\,.
\end{equation}
The energy is given by~\footnote{We have integrated by parts concerning the spatial derivative.}
\begin{equation}
\label{eq:energy-source}
    \begin{split}
        E&=\int\diff^3x\,\mathcal{H}\overset{\text{eom}}{=}\int\diff^3x\,\dfrac{1}{2}\left[(\dot{\phi})^2-\Ddot{\phi}\phi-J\phi\right]\,.
    \end{split}
\end{equation}
This form shows clearly that for the static case the energy depends directly on the source.

\paragraph{Massless case.}
Let us first consider the massless case.
To compute the energy we also need to know the solution of $\phi(x)$ for $r<R$ as a function of time. In fact in the static case ($t<0$) the Hamiltonian density is proportional to the charge density $J(x)$, which is only localized in $r<R$. Eq.~\eqref{eq:field_R} yields
\begin{equation}
    \phi(t<0,r\leq R) =\dfrac{1}{8\pi R}\left(3-\dfrac{r^2}{R^2}\right)\,.
\end{equation}
This result agrees with the one that can be derived by Gauss'Law.\\
\noindent

For the static case, i.e. for $t<0$, the energy of the field reads
\begin{equation}
    \begin{split}
        E&=\int\diff^3x\,\mathcal{H}=-\dfrac{1}{2}\int\diff^3x \phi J(x)=-\dfrac{4\pi}{2}\int\limits_0^R\diff{r}\, r^2\dfrac{g}{8\pi R}\left(3-\dfrac{r^2}{R^2}\right) \dfrac{3g}{4\pi R^3}\\
        &=-\dfrac{3g^2}{20\pi R}\,.
    \end{split}
\end{equation}

As the energy is conserved after the source is turned off, we can compute it for simplicity at $t\gg 2R$. We then get
\begin{align}
    &\int\diff^3x (\dot{\phi})^2= 4\pi\int\limits_{t-R}^{t+R}\diff{r}\, r^2 (\dot{\phi})^2=\dfrac{3g^2}{20\pi R}\,,\\
    &\int\diff^3x \ddot{\phi}\phi= 4\pi\int\limits_{t-R}^{t+R}\diff{r}\, r^2 \ddot{\phi}\phi=-\dfrac{3g^2}{20\pi R}\,,
\end{align}
so that
\begin{equation}
    E_{f,0}=\int\diff^3 x \dfrac{1}{2}\left[(\dot{\phi})^2- \ddot{\phi}\phi\right]=\dfrac{3g^2}{20\pi R}\,.
\end{equation}
The change of sign is consistent with the field being first bounded and then free propagating.

\paragraph{Massive case.}
In the massive case, we can again compute the energy for $t<0$:
\begin{equation}
\begin{split}
        E&=-\dfrac{4\pi}{2}\dfrac{3g}{4\pi R^3}\int\limits_0^R\diff{r}\, r^2 \dfrac{3g}{2\pi^2R^2 r}\int\limits_0^\infty \diff{k}\,\dfrac{1}{\omega_k^2}\dfrac{\sin(kr)}{k}\left[\dfrac{\sin(k R)}{kR}-\cos{(kR)}\right]\,,\\
        &=-\dfrac{9g^2}{(2\pi)^2 R^4}\int\limits_0^\infty \diff{k}\,\dfrac{1}{k^2 \omega_k^2}\left[\dfrac{\sin(k R)}{kR}-\cos{(kR)}\right]^2\,,\\
        &=-\dfrac{3g^2}{16\pi x^5 R}\left[(3-3x^2+2x^3)-3e^{-2x}(1+x)^2\right]\,,
        \end{split}
\end{equation}
where we have defined $x\equiv m_\phi R$\,. The above result consistently reproduces the massless case in the limit $m_\phi\to 0$.

\subsection{Starting from the Initial Field Configuration}
Another approach is to be agnostic about the coupling and matter configuration that sources the field. One can simply assume different field profiles at the moment of the disappearance of the source and let them evolve via the free KG equation.

For practical purposes, it is then convenient to trade the coupling of the field with the total energy of the initial field configuration, which is now promoted to a free parameter of the theory. 

\subsubsection{Energy of the Field}
The generic solution for the KG equation can be easily written using the Fourier transform
\begin{equation}
\label{eq_phi_x}
    \phi(x)=\dfrac{1}{(2\pi)^3}\int\limits \dfrac{\diff^3 k}{2\omega_k} \left[e^{i(\omega_k t-\Vec{k}\cdot \Vec{x})}\tilde{\phi}(k)+\text{h.c.}\right]=\dfrac{1}{(2\pi)^2 |\vec{x}|}\int\limits _0^\infty\diff k\dfrac{k\sin(k|\Vec{x}|)}{\omega_k} \left[e^{i\omega_k t}\tilde{\phi}(k)+\text{h.c.}\right]\,,
\end{equation}
where $\tilde{\phi}(k)$ is the Fourier Transform of $\phi(x)$. This also applies at $t=0$.
We then have
\begin{equation}
\label{eq:transform}
    \dfrac{\tilde\phi(k)}{\omega_k}=\int\diff^3x e^{i\vec{k}\cdot \Vec{x}}\phi(x,t=0)\,.
\end{equation}

In the absence of sources as applicable at $t>0$ the Hamiltonian density is given by
\begin{equation}
    \mathcal{H}=\dfrac{1}{2}\left[(\dot{\phi})^2+(\phi')^2+m^2\phi^2\right]\,.
\end{equation}
A short calculation yields the expected result,
\begin{equation}
E=      \dfrac{1}{2}\int\dfrac{\diff^3k}{(2\pi)^3}\, |\tilde\phi(k)|^2\,.
\end{equation}
This implies that per frequency
\begin{equation}
    \diff{E}(k)=\dfrac{1}{4\pi^2} k^2\,|\tilde\phi(k)|^2\,\diff{k}\,.
\end{equation}

As a cross-check of such a result, we take the result of Eq.~\eqref{eq:field_R} and compute the energy.
By matching to the form we need, we find
\begin{equation}
\label{eq:FT-easy}
    \tilde{\phi}(k)=\dfrac{3g}{ R^2}\dfrac{\mathcal{F}(kR)}{\omega_k k^2}\,.
\end{equation}
This yields
\begin{equation}
    E=\dfrac{3g^2}{16\pi x^5 R}\left[(3-3x^2+2x^3)-3e^{-2x}(1+x)^2\right]\,.
\end{equation}
This is consistent with the above results. 

\subsection{Time-Evolution and Average Forces}\label{app:average}
By combining Eqs.~\eqref{eq_phi_x}, \eqref{eq:transform}, one gets the generic time-evolution formula for a spherical symmetric field at $t>0$
\begin{equation}
    \phi(r,t)=\frac{4}{(2\pi)r}\int\limits_0^\infty \diff{k}\cos{(\omega_k t)}\sin(kr)\int\limits_{0}^\infty \diff{y}\, y\phi(y)\sin(ky)\,.
\end{equation}

For a massless field~\footnote{One can directly check that the solution satisfies the KG equation and that for $t=0$ we obtain
\begin{equation}
    \phi(r,0)=\phi(r)\,.
\end{equation}
}
\begin{equation}
    \begin{split}
        \phi(r,t)&=\dfrac{1}{2r}\int\limits_0^\infty\diff{x}\,x\phi(x)\left[\delta(x-r+t)+\delta(x-r-t)-\delta(x+r+t)-\delta(x+r-t)\right]\,,\\
        &=\dfrac{1}{2r}\left\{(r+t)\phi(r+t)+(r-t)\phi(r-t)\Theta(r-t)-(t-r)\phi(t-r)\Theta(t-r)\right\}\,.
    \end{split}
\end{equation}

Let us check some generic features and check some properties. We obtain
\begin{align}
    &\label{eq:1-time-app}\phi(r>t)=\dfrac{1}{2r}\left[(r+t)\phi(r+t)+(r-t)\phi(r-t)\right]\,,\\
    &\label{eq:2-time-app}\phi(t>r)=\dfrac{1}{2r}\left[(r+t)\phi(r+t)-(t-r)\phi(t-r)\right]\,.
\end{align}

As example, for $t>0$ and a YL-source we obtain, 
\begin{align}
\label{eq:KG-field}
    &\phi(r\leq R)=-\dfrac{g}{8\pi R^3}\times \begin{cases}
    \left(r^2-3R^2+3t^2\right)\,,& t\in [0,R-r]\\\
    \dfrac{1}{2r}(r-2R-t)(r+R-t)^2\,,&t\in[R-r,R+r]\\
    0\,,&t\in[R+r,+\infty)
    \end{cases}\,,\\
    &\phi(r\geq R)=\dfrac{g}{4\pi r}\times \begin{cases}
    1\,,& t\in [0,r-R]\\\
    -\dfrac{(r-2R-t)(r+R-t)^2}{4R^2}\,,&t\in[r-R,r+R]\\
    0\,,&t\in[r+R,+\infty)
    \end{cases}\,,
\end{align}
consistently with what previously found in Eqs.~\eqref{eq:field-r>R}, \eqref{eq:field-r<R}.

Finally, the force can be written as
\begin{align}
    &F(r>t)=\dfrac{t}{2r^2}\left[\phi(r+t)-\phi(r-t)\right]-\dfrac{1}{2r}\left[(r+t)\phi'(r+t)+(r-t)\phi'(r-t)\right]\,,\\
    &F(r<t)=\dfrac{t}{2r^2}\left[\phi(r+t)-\phi(t-r)\right]-\dfrac{1}{2r}\left[(r+t)\phi'(r+t)+(t-r)\phi'(t-r)\right]\,,
\end{align}
which allows for a more compact expression
\begin{equation}
    F(r,t)=\dfrac{t}{2r^2}\left[\phi(r+t)-\phi(|r-t|)\right]-\dfrac{1}{2r}\left[(r+t)\phi'(r+t)+|r-t|\phi'(|r-t|)\right]
\end{equation}

We might be interested in time-averaged quantities. We focus first on the field average and compare it then to the force. We will consider a signal that lasts within the time $t\in[r-R,r+R]$. 
We get
\begin{align}
     &\left<\phi(r>t)\right>=\dfrac{1}{2rR}\int\limits_{0}^R\diff{\tau}\,\left[(2r-\tau)\phi(2r-\tau)+\tau\,\phi(\tau)\right]\,,\\
     &\left<\phi(r<t)\right>=\dfrac{1}{2rR}\int\limits_{0}^R\diff{\tau}\,\left[(2r+\tau)\phi(2r+\tau)-\tau\,\phi(\tau)\right]\,,
\end{align}
It follows that if the initial field configuration has compact support, then the total average field is 0. Otherwise, one finds that, for $r\gg R$,
\begin{equation}
    \left<\phi\right>\approx\phi(2r)\,.
\end{equation}
For example, in the case of a static YL-field, one can check that
\begin{equation}
    \left<\phi\right>\sim\dfrac{1}{4\pi(2r)}=\dfrac{1}{8\pi r}\,,
\end{equation}
that matches explicit calculation.
One can similarly write the expressions for the average forces
\begin{align}
     &\label{eq:force-generic1}\left<F(r>t)\right>=\dfrac{1}{2rR}\int\limits_{0}^R\diff{\tau}\,\left\{\dfrac{r-\tau}{r}\left[\phi(2r-\tau)-\phi(\tau)\right]-\left[(2r-\tau)\phi'(2r-\tau)+\tau\phi'(\tau)\right]\right\}\,,\\
     &\label{eq:force-generic2}\left<F(r<t)\right>=\dfrac{1}{2rR}\int\limits_{0}^R\diff{\tau}\,\left\{\dfrac{r+\tau}{r}\left[\phi(2r+\tau)-\phi(\tau)\right]-\left[(2r+\tau)\phi'(2r+\tau)+\tau\phi'(\tau)\right]\right\}\,.
\end{align}
In the limit of $r\gg R$ we find that
\begin{equation}
\label{eq:average-force-general}
    \begin{split}
        \left<F(r>t)\right>\approx\left<F(r<t)\right>\approx&\dfrac{1}{2rR}\int\limits_{0}^R\diff{\tau}\,\left\{\left[\phi(2r)-\phi(\tau)\right]-\left[2r\phi'(2r)+\tau\phi'(\tau)\right]\right\}\,,\\
        &\approx\dfrac{1}{2rR}\int\limits_{0}^R\diff{\tau}\,\left\{\left[\phi(2r)-2r\phi'(2r)\right]-\left[\phi(\tau)+\tau\phi'(\tau)\right]\right\}\,,\\
        &\approx\dfrac{\phi(2r)-2r\phi'(2r)}{2r}-\dfrac{\phi(R)}{2rR}\approx -\dfrac{\phi(R)}{2r}\,.
    \end{split}
\end{equation}
This is the LO contribution if $\phi(R)\neq0$. This result matches the exact computations for both the YL- and CS-source. Notice that it is vanishing in the CS-case.

\footnotesize

\bibliographystyle{BiblioStyle}
\bibliography{Bibliography}
\end{document}